\documentclass[aps,twocolumn,superscriptaddress,showpacs]{revtex4}

\usepackage[dvips]{graphicx} 
\usepackage{color}           
\usepackage{amssymb}
\usepackage{bm}
\newcommand{\nc}{\newcommand}           
\nc{\vc}[1]     {\mbox{\boldmath $#1$}} 
\nc{\mapleft}[1]{                       
 \smash{\mathop{                        %
  \hbox to 0.90cm{\rightarrowfill} }\limits_{#1}}}

\nc{\red}[1]    {\textcolor{red}{#1}}  
\nc{\blue}[1]   {\textcolor{blue}{#1}}  
\nc{\green}[1]   {\textcolor{green}{#1}}  

\nc{\bras}[1]	{\langle #1 |}		
\nc{\kets}[1]	{|#1 \rangle}		

\nc{\mydraft}	{\setlength{\topmargin}{-1.5cm}}
\mydraft

\begin{document}

\title{Variational calculation of nuclear matter in finite particle number approach using unitary correlation operator and high-momentum pair methods}

\author{Takayuki Myo\footnote{takayuki.myo@oit.ac.jp}}
\affiliation{General Education, Faculty of Engineering, Osaka Institute of Technology, Osaka, Osaka 535-8585, Japan}
\affiliation{Research Center for Nuclear Physics (RCNP), Osaka University, Ibaraki, Osaka 567-0047, Japan}

\author{Hiroki Takemoto\footnote{takemoto@gly.oups.ac.jp}}
\affiliation{Osaka University of Pharmaceutical Sciences, Takatsuki, Osaka 569-1094, Japan} 

\author{Mengjiao Lyu\footnote{mengjiao@rcnp.osaka-u.ac.jp}}
\affiliation{Research Center for Nuclear Physics (RCNP), Osaka University, Ibaraki, Osaka 567-0047, Japan}

\author{Niu Wan\footnote{wanniu\_nju@163.com}}
\affiliation{School of Physics, Nanjing University, Nanjing 210093, China}

\author{Chang Xu\footnote{cxu@nju.edu.cn}}
\affiliation{School of Physics, Nanjing University, Nanjing 210093, China}

\author{Hiroshi Toki\footnote{toki@rcnp.osaka-u.ac.jp}}
\affiliation{Research Center for Nuclear Physics (RCNP), Osaka University, Ibaraki, Osaka 567-0047, Japan}

\author{Hisashi Horiuchi\footnote{horiuchi@rcnp.osaka-u.ac.jp}}
\affiliation{Research Center for Nuclear Physics (RCNP), Osaka University, Ibaraki, Osaka 567-0047, Japan}

\author{Taiichi Yamada\footnote{yamada@kanto-gakuin.ac.jp}}
\affiliation{College of Science and Engineering, Kanto Gakuin University, Yokohama 236-8501, Japan}

\author{Kiyomi Ikeda\footnote{k-ikeda@riken.jp}}
\affiliation{RIKEN Nishina Center, Wako, Saitama 351-0198, Japan}

\date{\today}

\begin{abstract}%
We propose a new variational method for describing nuclear matter from nucleon-nucleon interaction.
We use the unitary correlation operator method (UCOM) for central correlation to treat the short-range repulsion and further include the two-particle two-hole (2p2h) excitations of nucleon pair involving a large relative momentum, which is called ``high-momentum pair'' (HM).
We describe nuclear matter in finite size with finite particle number on periodic boundary condition
and increase the 2p2h configurations until we get the convergence of the total energy per particle.
We demonstrate the validity of this ``UCOM+HM'' framework by applying it to the symmetric nuclear and neutron matters with the Argonne V4$^\prime$ potential having short-range repulsion.
The nuclear equations of state obtained in UCOM+HM are fairly consistent to those of other calculations such as Brueckner-Hartree-Fock and 
auxiliary field diffusion Monte Carlo in the overall density region.
\end{abstract}

\pacs{
21.65.-f  
}
\maketitle 

\section{Introduction} \label{sec:intro}

The nucleon-nucleon ($NN$) interaction is fundamental in the descriptions of finite nuclei and  nuclear matter.
The $NN$ interaction shows a strong repulsion at short-range part and a strong tensor force at intermediate-range and long-range parts \cite{pieper01,wiringa95,wiringa02,feldmeier98,neff04}.
The above two characteristics of the $NN$ interaction produce the high-momentum components in nucleon motion for nuclear system. 
The short-range repulsion decreases the amplitudes of a nucleon pair with short distance.
The tensor force provides the $D$-wave state of a nucleon pair due to the strong $S$-$D$ coupling.

For finite nuclei, we have described the short-range and tensor correlations with several kinds of theoretical frameworks.
One is the shell model-type method, which is called the ''tensor-optimized shell model'' (TOSM) \cite{myo05,myo09,myo11,myo12}.
In TOSM, we can optimize the two-particle two-hole (2p2h) states without truncation of the particle states.
The 2p2h excitations express the strong tensor correlation in finite nuclei.
In order to treat the short-range correlation, we combine TOSM with the central part of the unitary correlation operator method, which we call ``central-UCOM'' \cite{feldmeier98,neff04,myo09}. 
In the central-UCOM, the central-type unitary operator is considered to decrease the amplitudes of a nucleon pair at short-range distance.
In UCOM, the transformed Hamiltonian is taken within two-body operators, while the exact transformation emerges many-body operators.
This two-body truncation works reasonably in the description of the short-range correlation \cite{feldmeier98}.
In TOSM+UCOM, we explicitly treat the short-range and tensor correlations in nuclei.
As a result, the shell model-like states can be described nicely with the correct order of the energy levels of $p$-shell nuclei \cite{myo11,myo12,myo14}.

There are other approaches to describe the $NN$ correlation explicitly in finite nuclei.
We have developed a new variational theory for finite nuclei \cite{myo15,myo17a,myo17b,myo17c,myo17d,myo18,lyu18a,lyu18b}.
We use the antisymmetrized molecular dynamics (AMD) \cite{kanada03} as the basis wave function
and employ two kinds of the correlation functions of the central-operator type and the tensor-operator type to describe the $NN$ correlations.
These correlation functions are successively multiplied to the AMD wave function.
We name this method the ''tensor-optimized antisymmetrized molecular dynamics'' (TOAMD) \cite{myo15}.
In the previous works for $s$-shell nuclei \cite{myo17a,myo17b,myo17c,myo17d}, we have shown that TOAMD can reproduce the results of Green's function Monte Carlo (GFMC) with the bare $NN$ interaction.

There is another approach, which is recently developed to treat the high-momentum components from the $NN$ interaction in finite nuclei explicitly.
We adopt AMD as the basis state again, and superpose the AMD basis states, in which the nucleon pair involves a large relative momentum.
We call this nucleon pair ``high-momentum pair'' (HM) and name this method ``high-momentum AMD'' (HM-AMD) \cite{myo18,lyu18a,lyu18b,myo17e,zhao18}.
In HM-AMD, we successfully treat the high-momentum components in finite nuclei arising from the $NN$ interaction.
It is noted that the above high-momentum pair produces the large transfer momentum in the nucleon pair with an opposite direction to each other \cite{itagaki18}.
This feature is also included in the correlation functions in TOAMD \cite{myo17d,lyu18b}.

In the study of nuclear matter, the $NN$ interaction is an important ingredient and a decisive role on the description of the nuclear equations of state of symmetric nuclear and neutron matters, information of which is closely related to the neutron star physics.
There are several theories of nuclear matter treating the $NN$ correlations, such as Brueckner-Hartree-Fock (BHF) and Brueckner-Bethe-Goldstone (BBG) \cite{baldo12}, Fermi hypernetted chain \cite{lovato11}, Green's function Monte Carlo (GFMC) \cite{carlson03,carlson15}, auxiliary field diffusion Monte Carlo (AFDMC) \cite{sarsa03,gandolfi09,gandolfi14}, and coupled cluster theory (CC) \cite{hagen14,LNP936}.

Baldo et al. have performed the benchmark calculations of nuclear matter based on these different theories with the same $NN$ interactions \cite{baldo12}.
From the results, it is found that these theories provides the similar energies per particle of nuclear matter to each other in case of the short-range correlation, but, there exists a large difference between them for the $NN$ interaction including tensor and $LS$ forces. In particular, treatment of tensor force as tensor correlation becomes a key for this difference. Tensor correlation is an intermediate and long-range correlation and this property induces many-body correlations in nuclear matter. Tensor correlation also plays an important role on the saturation property of the equation of state for symmetric nuclear matter.
It should be clarified to treat the tensor correlation consistently in nuclear matter in addition to the short-range correlation from the bare $NN$ interaction.

In other groups, Hu et al. have performed the calculations of nuclear matter using Hartree-Fock approximation of Fermi sphere with the central-UCOM in relativistic and non-relativistic frameworks \cite{hu10,hu12}. They start from the Bonn potential and the results shows that the short-range correlation is nicely treated using the central-UCOM in the Hartree-Fock level.
Yamada has recently proposed a new variational theory, which is called ``tensor-optimized Fermi sphere'' (TOFS) \cite{yamada18a,yamada18b}. In TOFS, the correlation functions and their multiple products are multiplied to the Fermi-sphere state,
and the correlation functions are treated as variational functions to be determined in the total energy minimization.
This concept is the same as that of TOAMD for finite nuclei as explained before.
Recently, he and his collaborators have applied TOFS to describe the symmetric nuclear matter with Argonne V4$^\prime$ (AV4$^\prime$) central potential having short-range repulsion \cite{wiringa02}.
The equation of state successfully reproduces the benchmark calculations obtained in other theories.

In this study, we propose a new variational method for nuclear matter treating the $NN$ interaction directly
in a different method of TOFS.
We explicitly include many-body correlations from the $NN$ interaction in the nuclear matter wave function 
and for this purpose, we describe the nuclear matter in a finite size box on periodic boundary condition
and adopt the finite particle number approach. This is treated in the same manner as used in GFMC, AFDMC and CC and also in the electron systems \cite{lin01}.
This method is capable of including the particle-hole excitations in the wave function
in a similar way of finite nuclei.
We further use the central-UCOM to treat the short-range correlation and perform the configuration mixing of 2p2h excitations.
This approach is similar to our TOSM+UCOM method for finite nuclei and regarded as the extension of Hu's work
beyond the Hartree-Fock approximation.
It is interesting to investigate the 2p2h effect in addition to UCOM for nuclear matter.
The 2p2h excitations occur for the nucleon pair in nuclear matter and this nucleon pair receives 
the transfer momentum in the opposite directions to each other, which induces a large relative momentum in a nucleon pair. This property is also utilized in HM-AMD and TOAMD for finite nuclei.
It is noted that the 2p2h excitations are important to describe the tensor correlation.

In this paper, we explain the details of this new framework and focus on the description of the short-range correlation coming from the $NN$ short-range repulsion.
We investigate the applicability of the present method to the symmetric nuclear and neutron matters.
We use the Argonne V4$^\prime$ central potential, which is renormalized from the realistic AV18 potential \cite{wiringa02}.
We compare our results with those of other theories as benchmark calculations and confirm the validity of the present method.

In Sec.~\ref{sec:method}, we explain the nuclear matter wave function within the 2p2h excitations,
and also UCOM for short-range correlation.
In Sec.~\ref{sec:result}, we present the results of symmetric nuclear and neutron matters.
A summary is given in Sec.~\ref{sec:summary}.

\section{Methods}\label{sec:method}

\subsection{Wave function}
\begin{figure}[t]
\begin{center}
\includegraphics[width=6.0cm,clip]{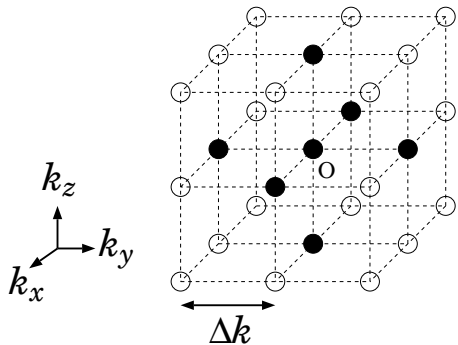}
\caption{Lattice in momentum space with Cartesian representation.
The configuration of the occupied states with the number of grid points $N_g=7$ is shown using solid circles.
Open circles indicate the unoccupied states. The grid spacing is $\Delta k$.}
\label{fig:cube}
\end{center}
\end{figure}

We describe nuclear matter in a finite size of cubic box with a finite mass number $A$ under the periodic boundary condition.
The basis wave function of nuclear matter is given in the form of the Slater-determinant.
We first define the 0p0h state of nuclear matter as
\begin{eqnarray}
    \kets{{\rm 0p0h}}
&=& \frac{1}{\sqrt{A!}} {\rm det}\left\{ \prod_{i=1}^A \phi_{\alpha_i}(\vc{r}_i) \right\} , 
    \label{eq:0p0h}
    \\
    \phi_\alpha(\vc{r})
&=& \frac{1}{\sqrt{L^3}} e^{i \bm{k}_\alpha\cdot\bm{r}} \chi^\sigma_\alpha \chi^\tau_\alpha ,
    \\
    \langle \phi_\alpha|\phi_{\alpha'} \rangle
&=& \delta_{\alpha,\alpha'}.
\end{eqnarray}
Here $\phi_\alpha(\vc{r})$ is a nucleon wave function in a plane wave form with a momentum (wave number) $\vc{k}_\alpha$, and $\chi^\sigma_\alpha$ and $\chi^\tau_\alpha$ are the spin and isospin component, respectively.
In this study, $\chi^\sigma_\alpha$ is the up or down component and $\chi^\tau_\alpha$ is a proton or a neutron.
The index $\alpha$ is a representative quantum number for momentum, spin and isospin.
The side of cube $L$ is related to the normalization of the nucleon wave function. 
We employ the periodic boundary condition such as $\phi_\alpha(\vc{r}+L\vc{\hat x})=\phi_\alpha(\vc{r})$ in the  Cartesian coordinate system.
From this condition momentum is discretized with the gap $\displaystyle \Delta k=\frac{2\pi}{L}$.
We consider a lattice in momentum space as shown in Fig. \ref{fig:cube}, where each grid point represents the nucleon state with the eigenstate of momentum.
Using the integer vector $\vc{n}=(n_x, n_y, n_z)$, we represent the momentum of each nucleon on a lattice as $\displaystyle \vc{k}=\frac{2\pi}{L}\vc{n}$.

From the symmetry of the total wave function of nuclear matter in momentum space, we choose the particle number as the magic number corresponding to the shell closure on a lattice in momentum space, which is the number of grid points $N_g=1, 7, 19, 27, 33, 57, 81, \cdots$, occupied from the center of lattice in the order of the smaller momenta.
The case of $N_g=7$ is shown with the Cartesian representation of momentum in Fig. \ref{fig:cube}.
We prepare the single nucleon wave function with the momentum $\vc{k}_i$ with $i=1,\cdots,N_g$ to construct the 0p0h state, $\kets{{\rm 0p0h}}$ in Eq.~(\ref{eq:0p0h}).
For each grid number $N_g$, considering spin-isospin quantum numbers, the total particle number becomes a mass number $A=4N_g$ for symmetric nuclear matter and also a neutron number $N=2N_g$ for neutron matter.

In the calculation of symmetric nuclear matter, we fix the mass number $A$ as magic number and give the density $\rho$, which define the volume $\displaystyle V=\frac{A}{\rho}$ for cube and the side of cube $L=V^{1/3}$. For neutron matter, we follow the same prescription with a neutron number $N$.
Under these conditions we calculate the kinetic and potential energies for the 0p0h state.
This 0p0h state corresponds to the Hartree-Fock state of Fermi sphere for the infinite matter with the limit of $L=\infty$,
where the Fermi momentum $k_F$ can be defined as $k_F=(3\pi^2\rho/2)^{1/3}$ for symmetric nuclear matter,
and $k_F=(3\pi^2\rho)^{1/3}$ for neutron matter.

So far, there are several calculations of nuclear matter with finite size and finite particle number, such as GFMC \cite{carlson03}, AFDMC \cite{sarsa03,gandolfi09} ,and CC \cite{hagen14}.
In these calculations, it is shown that in the region of smaller magic numbers, the numerical results with $N_g=33$ provide the kinetic and potential energies per particle close to those of the Hartree-Fock infinite matter. 
In the present study, we investigate the particle-number dependence of the solutions in the present method.

In the calculation of potential matrix elements in a cubic box, we include the tail effect to take the contribution from the neighboring boxes similarly to AFDMC \cite{sarsa03} and CC \cite{LNP936};
the interparticle motion for the potential coordinate is integrated out with infinite space. 
This is also considered for the two-body operators in UCOM, as is explained later.

Next, we include the 2p2h configurations as follows.
Considering the symmetric nuclear matter, we assign the states with index $i$ and $j$ being the hole states with $i,j=1,\cdots,A=4N_g$,
which occupy the states from lower magnitude of momenta. 
We define the particle states using the index $m$ and $n$ with $m,n>A$.
Hence the 2p2h states are written as 
\begin{eqnarray}
\kets{{\rm 2p2h}}
&=& \kets{mn;i^{-1} j^{-1}}.
\label{eq:2p2h}
\end{eqnarray}
In the 2p2h states, the following momentum conservation is held between two-hole and two-particle states;
\begin{eqnarray}
\vc{k}_i+\vc{k}_j &=& \vc{k}_m+\vc{k}_n.
\label{eq:momentum}
\end{eqnarray}
Here we introduce the transfer momentum $\vc{q}$, which affects the relative momentum of two nucleons
in the particle states as
\begin{eqnarray}
\vc{k}_m &=& \vc{k}_i + \vc{q},
\\
\vc{k}_n &=& \vc{k}_j - \vc{q}.
\end{eqnarray}
This momentum $\vc{q}$ is also discretized from the periodic boundary condition
and defined as $\displaystyle\vc{q}=\frac{2\pi}{L}\vc{n}_q$.
The integer vector $\vc{n}_q=(n_{q\,x}, n_{q\,y}, n_{q\,z})$ represents the mode of transfer momentum and we put a maximum integer on its magnitude as $n_q^{\rm max}$ where $n_q^{\rm max}\geq|\vc{n}_q|$.
This $n_q^{\rm max}$ controls the number of 2p2h configurations and the model space of the present calculation.
For spin-isospin components of the 2p2h states, the $z$-components of spin and isospin are conserved between two-hole and two-particle states
in addition to the momentum given in Eq.~(\ref{eq:momentum}).
We take all of the available 2p2h configurations under three conditions,
which include the exchange of spin-isospin components between two nucleons caused by the $NN$ interaction.
We check the convergence of the present solutions by increasing $n_q^{\rm max}$.

Physically, the above 2p2h states can bring the high-momentum components in nuclear matter when the transfer momentum $\vc{q}$ is sufficiently large.
This high-momentum component is important to describe the short-range correlation and tensor correlation in nuclear matter as well as in finite nuclei.
We use the same approach to introduce the various momentum component in nuclear matter as done in finite nuclei with HM-AMD \cite{myo17e,lyu18a,lyu18b,zhao18}, and 
call the nucleon pair with high-momentum components as ``high-momentum pair'' (HM) hereafter.
We take the large value of $\vc{q}$ as much as possible to converge the results.
We can simply estimate the effect of high-momentum component as follows;
For example, when we take $A=132$ and $\rho=0.17$ fm$^{-3}$ as normal density, the side of cube is $L=9.2$ fm and the momentum gap is 
$\Delta k=0.68$ fm$^{-1}$. If we take $n_q^{\rm max}=7$, the magnitude of maximum transfer momentum $|\vc{q}|=4.8$ fm$^{-1}$, which is more than three times of the empirical Fermi momentum $k_F=1.4$ fm$^{-1}$, and sufficient to shift the nucleon motion to the intermediate- and high-momentum regions in the nuclear matter.

\begin{figure}[th]
\begin{center}
\includegraphics[width=7.5cm,clip]{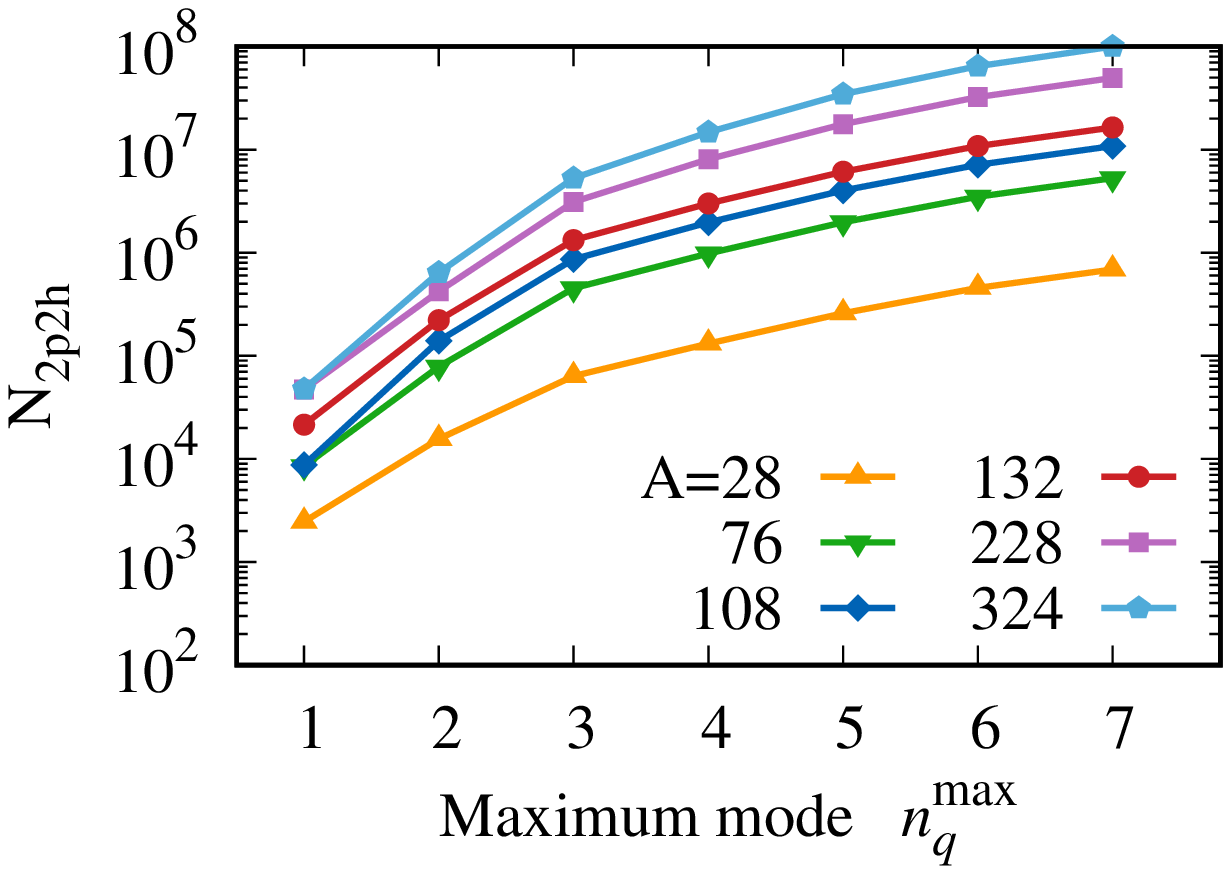}
\includegraphics[width=7.5cm,clip]{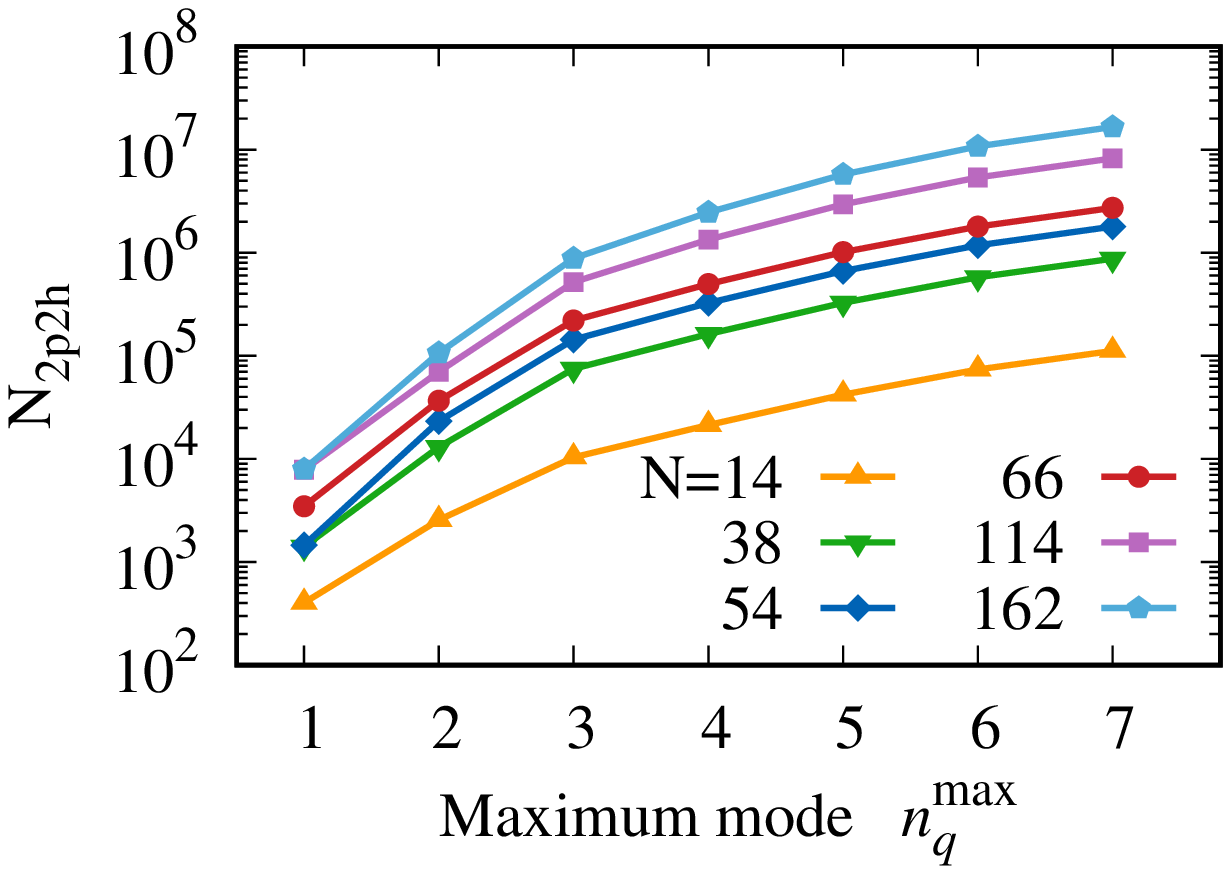}
\caption{Numbers of the 2p2h configurations $N_{\rm 2p2h}$ in symmetric nuclear matter (upper panel) and neutron matter (lower panel)
with six kinds of magic numbers as functions of maximum mode of transfer momentum $n_q^{\rm max}$.}
\label{fig:num_2p2h}
\end{center}
\end{figure}

Finally, we give the total wave function of nuclear matter in the space of the 0p0h+2p2h configurations 
and explicitly write it in the linear combination form as
\begin{eqnarray}
\Phi
&=& C_0\kets{{\rm 0p0h}} + \sum_{p=1}^{N_{\rm 2p2h}} C_p \kets{{\rm 2p2h},p},
\label{eq:WF}
\end{eqnarray}
where we allow the excitations from all kinds of pairs in the hole states.
The index $p$ is to distinguish the 2p2h configurations with indices $i,j,m,$ and $n$ in Eq.~(\ref{eq:2p2h}).  
The 0p0h state is assigned as $p=0$.
The number of the 2p2h configurations is $N_{\rm 2p2h}$.
In Fig.~\ref{fig:num_2p2h}, we show the values of $N_{\rm 2p2h}$ in symmetric nuclear matter and neutron matter with six kinds of magic numbers as functions of maximum mode of transfer momentum $n_q^{\rm max}$.
The coefficient $C_p$ is the amplitude of each basis state and determined variationally.
We perform the energy variation of nuclear matter with these basis states,
which results in solving the eigenvalue problem of the Hamiltonian matrix with the $N_{\rm 2p2h}+1$ dimension.
This is carried out by using the power method 
and we obtain the ground-state energy $E$ and the configuration amplitudes $\{C_p\}$ as an eigenvector.
Finally we discuss the energy per particle $E/A$ and the corresponding Hamiltonian components.
We use the same method for neutron matter.

\subsection{Unitary correlation operator method (UCOM)}

We use the central-UCOM to describe the short-range correlation in nuclear matter as well as the 2p2h excitations. This combination increases many-body correlations more in nuclear matter.
We explain the central-UCOM as a central-type correlation, which can soften the short-range repulsion in the $NN$ interaction \cite{feldmeier98,neff04}. 
We define the unitary operator $C_r$ using the pair-type Hermite generator $g$ as
\begin{eqnarray}
C_r   &=&\exp\left(-i\sum_{i<j} g_{ij}\right)
~=~ \prod_{i<j}^A c_{r,ij},
\label{eq:ucom}
\end{eqnarray}
where $c_{r,ij}$ is the operator for one pair in the $A$-body system.
We consider the correlated wave function $\Psi$ involving the short-range correlation,
which is expressed using the uncorrelated wave function $\Phi$ and the operator $C_r$ as $\Psi=C_r\Phi$. 
The transformed Schr\"odinger equation for $\Phi$ is written as $\tilde H \Phi=E\Phi$, in which the transformed Hamiltonian is defined as $\tilde H = C_r^\dagger H C_r$. 
In principle, the operator $C_r$ is a many-body operator and the transformed Hamiltonian $\tilde H$ can be a many-body operator.
In case of the short-range correlation, it is reasonable to truncate $\tilde H$ at the level of two-body operators \cite{feldmeier98,myo17b}.

The explicit form of the operator $g$ in Eq.~(\ref{eq:ucom}) is defined as
\begin{eqnarray}
g &=& \frac12 \bigl\{ p_r s(r)+s(r)p_r\bigr\} ,
\label{eq:ucom_g}
\end{eqnarray}
where the operator $p_r$ is the component of the relative momentum parallel to the relative coordinate between nucleons.
The function $s(r)$ represents the shift of the relative wave function at the relative distance $r$. 
In the calculation of central-UCOM, the function $R_+(r)$ is employed instead of $s(r)$ \cite{feldmeier98},
and this $R_+(r)$ represents the transformed distance of the original one $r$, and has a relation to $s(r)$ as 
\begin{eqnarray}
\frac{dR_+(r)}{dr}&=& \frac{s [ R_+(r) ]}{s(r)} , 
\\
c_r^\dagger r c_r &=& R_+(r).
\end{eqnarray}
The function $R_+(r)$ can decrease the amplitude of the relative wave function at short-range part as the short-range correlation. 
The transformations of other operators are given in details in Refs. \cite{feldmeier98,neff04}. 
The Hamiltonian, $H=T+V$ with two-body potential $V$ for $A$-body system is transformed as
\begin{eqnarray}
  \tilde H
  &=& C_r^\dagger T C_r
+C_r^\dagger V C_r
~=~ \tilde T + \tilde V,
\\
\tilde T
&=& \sum_{i=1}^A t_i
+ \sum_{i<j}^A u_{ij},
\qquad
\tilde V
~=~ \sum_{i<j}^A \tilde v_{ij}.
\label{eq:ham}
\end{eqnarray}
The transformed kinetic energy operator $\tilde T$ consists of the uncorrelated one-body term $t_i$ and the correlated two-body term $u_{ij}$. The latter term $u_{ij}$ depends on the momentum and angular momentum of the relative motion \cite{feldmeier98,neff04}.
In the transformed potential $\tilde V$, the radial component in $\tilde v$ is given as $v(R_+(r))$, which is transformed from the original $v(r)$ at the relative distance $r$.

The functional form of $R_+(r)$ is determined in the energy variation of the total system.
e parametrize $R_+(r)$ for the even channel with positive parity and for the odd channel with negative parity in the same manner used in Refs. \cite{feldmeier98,neff04} as
\begin{eqnarray}
	R_{+}^{\rm even}(r)
&=&	r + \alpha \left(\frac{r}{\beta}\right)^\gamma \exp\left[-\exp\left(\frac{r}{\beta}\right)\right] , 
        \label{eq:R+_even}
        \\
	R_{+}^{\rm odd}(r)
&=&	r + \alpha \left(1-\exp\left[-\frac{r}{\gamma}\right]\right)
\nonumber
\\
&\times&
 \exp\left[-\exp\left(\frac{r}{\beta}\right)\right] , 
        \label{eq:R+_odd}
\end{eqnarray}
where the parameters $\alpha$, $\beta$, and $\gamma$ are determined variationally.

\begin{table}[b]
\begin{center}
\caption{Three parameters in $R_+(r)$ of UCOM for two kinds of even channels with singlet ($^1E$) and triplet ($^3E$) states in $^4$He.}
\label{tab:R+} 
\begin{tabular}{c|ccccc}
\noalign{\hrule height 0.5pt}
         & $\alpha$ &  $\beta$ & $\gamma$ \\
\noalign{\hrule height 0.5pt}
$^1E$~   &~1.40~  &~ 1.02~ & ~0.33~ \\
$^3E$~   &~1.32~  &~ 0.94~ & ~0.42~ \\
\noalign{\hrule height 0.5pt}
\end{tabular}
\end{center}
\end{table}

We demonstrate the effect of UCOM with the AV4$^\prime$ potential in finite nuclei, $^4$He.
In this calculation, we choose the uncorrelated wave function $\Phi$ of $^4$He as the $(0s)^4$ configurations of the harmonic oscillator (HO) basis state with the length parameter $b$.
Hence the correlated wave function $\Psi$ depends on the length $b$.
We determine the length $b$ and $\alpha$, $\beta$, and $\gamma$ in $R_+(r)$ in the energy minimization of $^4$He with the AV4$^\prime$ and point Coulomb potentials.
In Table \ref{tab:R+}, we list the parameters in UCOM for even channel of $^4$He
and the length $b$ is optimized as 1.19 fm at the ground state.

\begin{figure}[t]
\begin{center}
\includegraphics[width=7.5cm,clip]{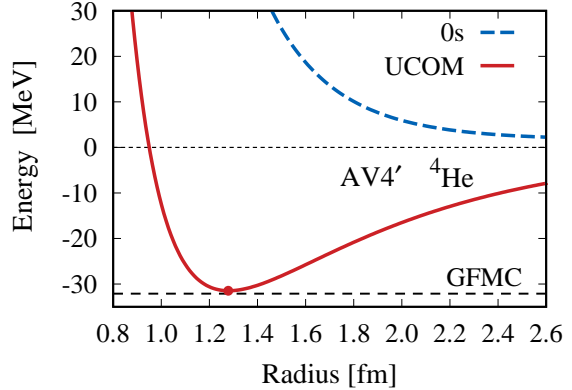}
\caption{Energy surface of $^4$He with AV4$^\prime$ potential as functions of radius.
The units of energy and radius are MeV and fm, respectively. 
The red-solid (blue-dashed) line is the results with (without) UCOM. The dashed line is the results of GFMC \cite{wiringa02}.}
\label{fig:4He}
\end{center}
\end{figure}

We show the results of $^4$He in Fig.~\ref{fig:4He} by changing the parameter of $b$ in HO, which affects the radius. 
We show two lines, which are the total energies with and without UCOM to represent the effect of UCOM.
The calculation with UCOM gains the energy largely from the uncorrelated $0s$ state
and reproduces well the energy of GFMC at the energy minimum point.
This result indicates that UCOM nicely works to treat the short-range repulsion in $NN$ interaction.

For the nuclear matter calculation, we determine the parameters of $R_+(r)$ in UCOM, again.
In Tables \ref{tab:R+_SNM} and \ref{tab:R+_NM}, we list these parameters for symmetric nuclear matter and neutron matter, respectively.
We obtain these parameters by minimizing the energy of the 0p0h state of nuclear matter with UCOM, corresponding to the 0p0h+UCOM calculation. In this calculation, we use the normal density $\rho=0.17$ fm$^{-3}$ and choose the magic number of grid point $N_g=33$ on a lattice in momentum space, leading to $A=132$ for symmetric nuclear matter and $N=66$ for neutron matter. These numbers are known to simulate well the infinite matter property under the Hartree-Fock approximation \cite{gandolfi09,hagen14}.
We tried the calculations with other particle numbers, and have confirmed that the differences of the  parameters in $R_+(r)$ for UCOM are negligible.
We also calculate the Hartree-Fock infinite matter with UCOM using the same $R_+(r)$.

\begin{table}[b]
\begin{center}
\caption{Three parameters in $R_+(r)$ of UCOM in four channels of symmetric nuclear matter with a mass number $A=132$.}
\label{tab:R+_SNM} 
\begin{tabular}{c|cccc}
\noalign{\hrule height 0.5pt}
              & $\alpha$ &  $\beta$ & $\gamma$ \\
\noalign{\hrule height 0.5pt}
$^1E$~         &~ 1.36~  &~ 0.98~  & ~0.33~ \\
$^3E$~         &~ 1.24~  &~ 0.94~  & ~0.39~ \\
$^1O$~         &~ 1.50~  &~ 1.26~  & ~0.87~ \\
$^3O$~         &~ 0.69~  &~ 1.39~  & ~0.28~ \\
\noalign{\hrule height 0.5pt}
\end{tabular}
\end{center}
\end{table}
\begin{table}[t]
\begin{center}
\caption{Three parameters in $R_+(r)$ of UCOM in two channels of neutron matter with a mass number $N=66$.}
\label{tab:R+_NM} 
\begin{tabular}{c|cccc}
\noalign{\hrule height 0.5pt}
              & $\alpha$ &  $\beta$ & $\gamma$ \\
\noalign{\hrule height 0.5pt}
$^1E$~         &~1.33~  &~ 1.00~ & ~0.31~ \\
$^3O$~         &~0.65~  &~ 1.39~ & ~0.24~ \\
\noalign{\hrule height 0.5pt}
\end{tabular}
\end{center}
\end{table}

\begin{figure}[t]
\begin{center}
\includegraphics[width=7.5cm,clip]{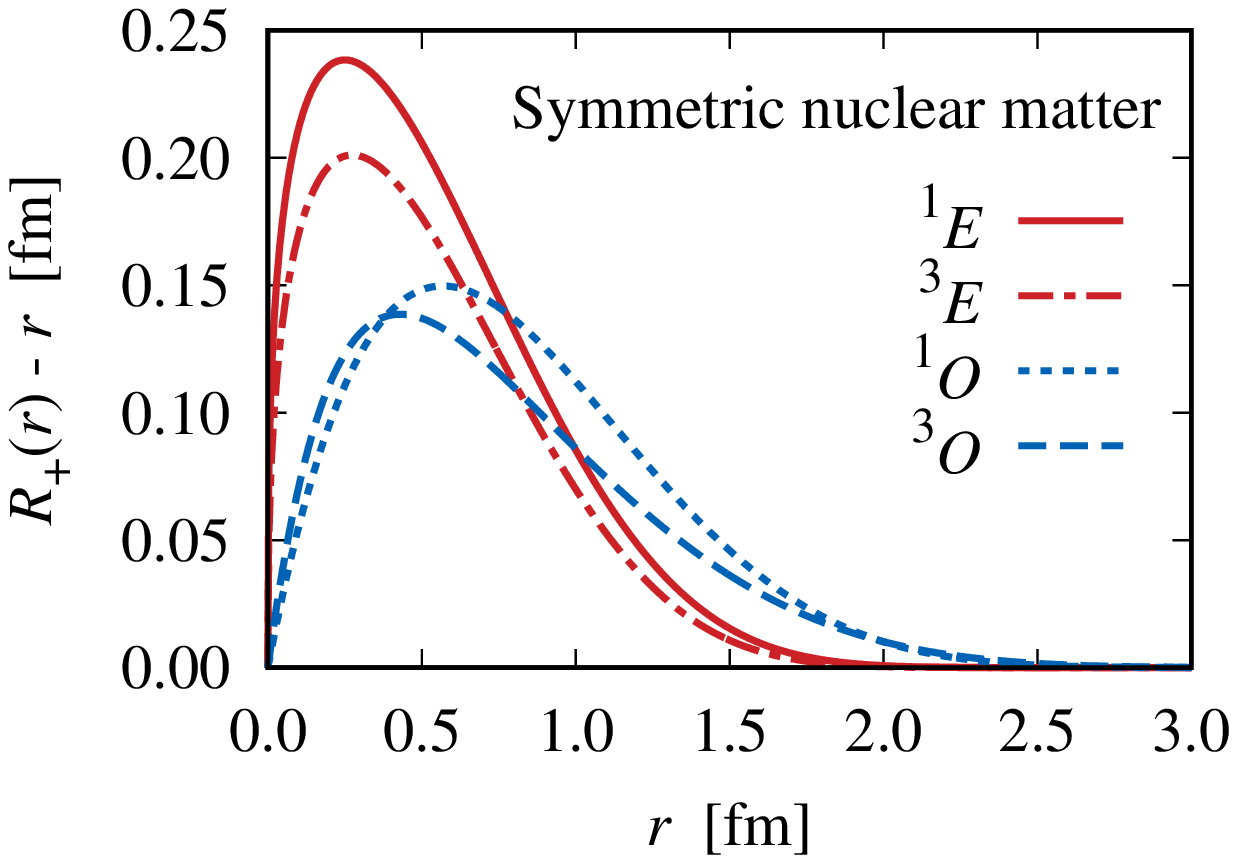}
\includegraphics[width=7.5cm,clip]{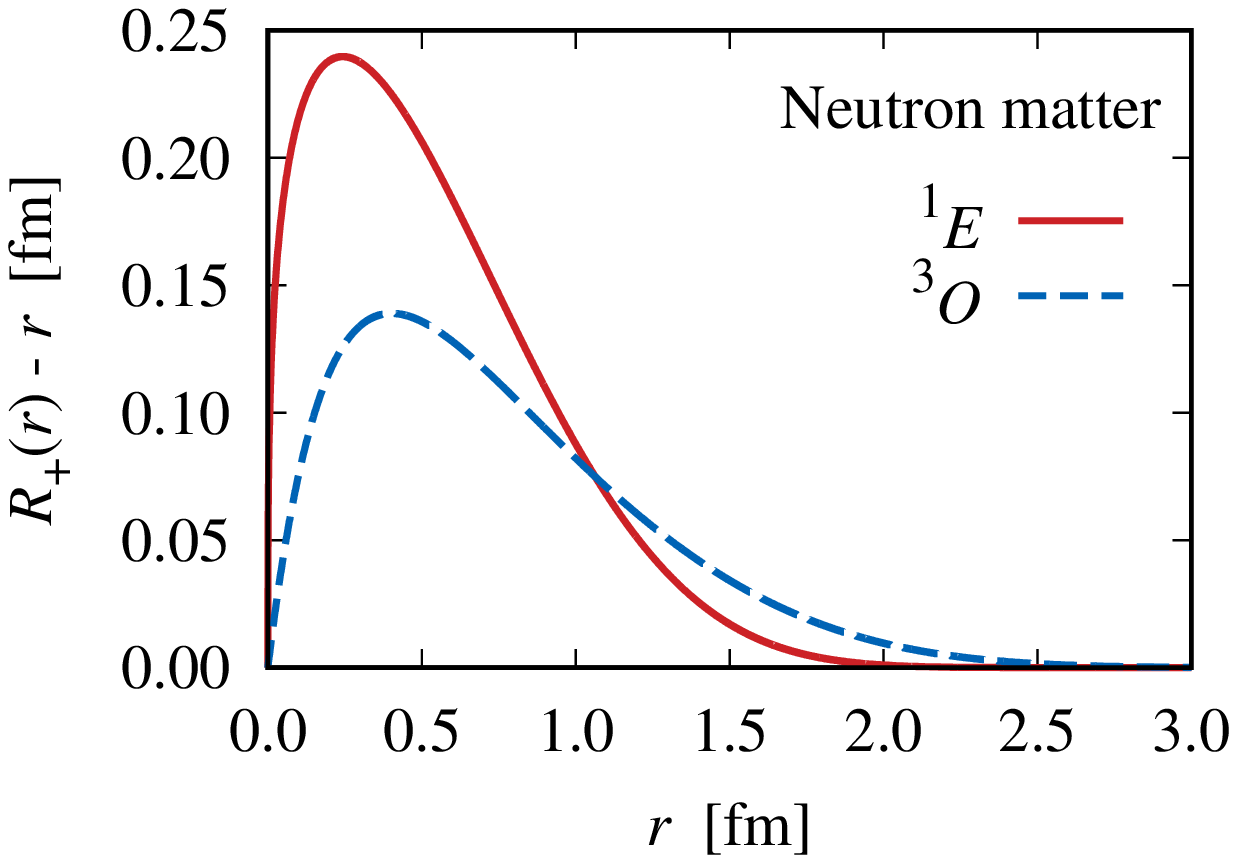}
\caption{Shift functions $R_+(r)$ in UCOM for each channel in symmetric nuclear matter (upper panel) and neutron matter (lower panel), respectively.
The differences between $R_+(r)$ and the original distance $r$ are shown as the shift of the wave function.}
\label{fig:R+}
\end{center}
\end{figure}

In Fig.~\ref{fig:R+}, we plot the distribution of $R_+(r)-r$ as the difference between the transformed and the original distances in a nucleon pair, 
which corresponds to the shift of the relative wave function at the distance $r$.
The distributions for symmetric nuclear matter and neutron matter show the similar shapes,
and the even channel tends to give a large amount of shift;
maximally the even channel gives around 0.20 -- 0.25 fm of the shift at about 0.25 fm of the relative distance.
The odd channel gives 0.13 -- 0.15 fm of the shift at about 0.5 fm of the relative distance.
We use these parameters in UCOM throughout the present calculations of nuclear matter.

Physically, the UCOM transformation induces the two-body correlation in many-body wave function.
In the present wave function of nuclear matter, we include up to the 2p2h excitations with high-momentum components in the basis states in Eq.~(\ref{eq:WF}).
This indicates that totally the 4p4h correlations are treated in the present approach.
We name the present method as ``UCOM+HM'' hereafter.

\section{Results}\label{sec:result}

\subsection{Symmetric nuclear matter}\label{sec:SNM}

\begin{figure}[b]
\begin{center}
\includegraphics[width=7.5cm,clip]{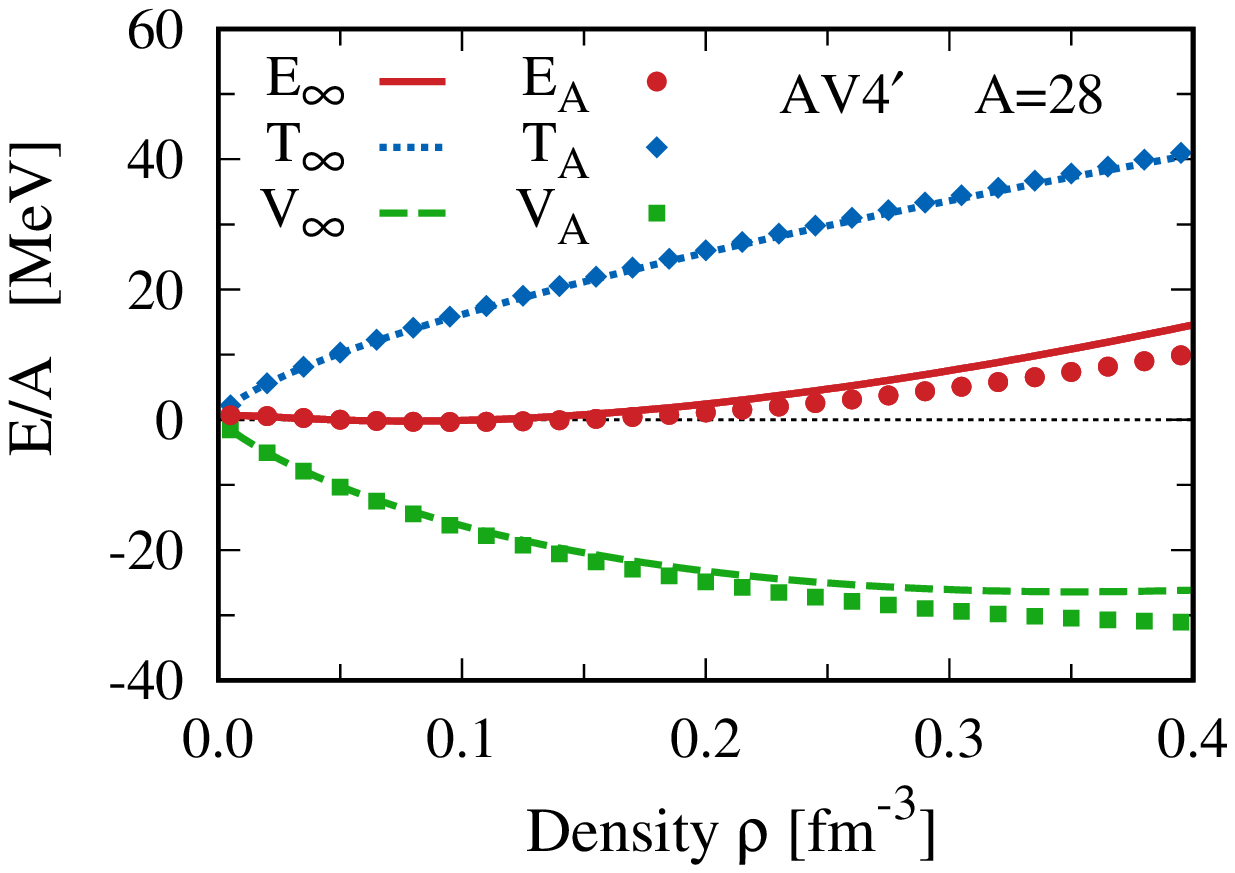}
\includegraphics[width=7.5cm,clip]{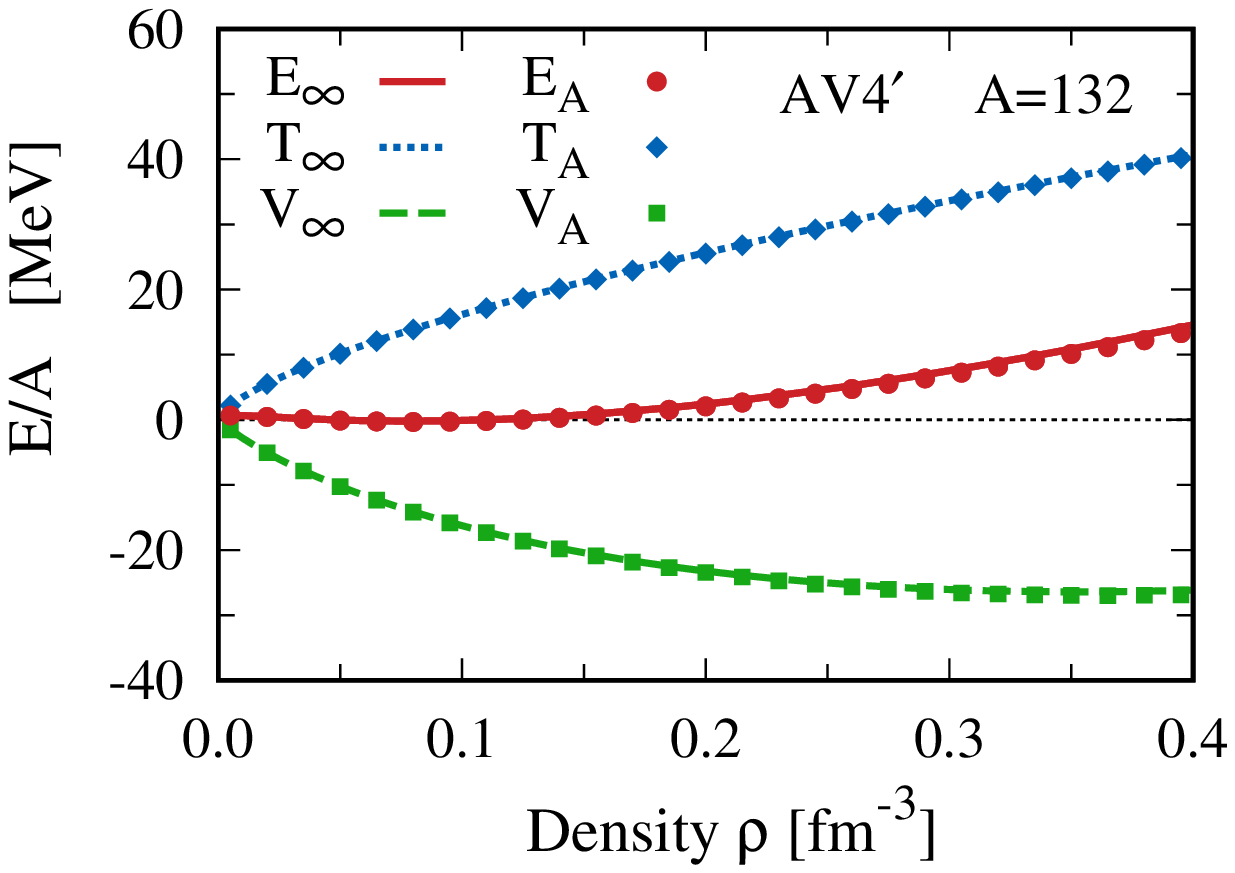}
\caption{Hamiltonian components per particle for the 0p0h state of symmetric nuclear matter without UCOM as functions of density $\rho$ with dots.
Mass number is $A=28$ (upper panel ) and $A=132$ (lower panel). 
Dots with red-circle, blue-diamond and green-square indicate the total ($E$), kinetic ($T$), and potential ($V$) energies, respectively. 
Results are compared with the Hartree-Fock solutions of Fermi sphere for infinite size with lines.
}
\label{fig:HF_SNM}
\end{center}
\end{figure}

We discuss the symmetric nuclear matter with UCOM+HM.
First we show two kinds of the results of the 0p0h and 0p0h+UCOM calculations with finite size and finite mass numbers. We compare them with those of infinite matter with infinite mass number to confirm the applicability of the present finite size approach. 
In Fig.~\ref{fig:HF_SNM}, we show the energy per particle as functions of density $\rho$ for mass number $A$
with only the 0p0h configuration without UCOM, which corresponds to the Hartree-Fock (HF) state of Fermi sphere for infinite matter.
We show the typical two cases of mass numbers $A=28$ and $132$ in addition to the HF solutions with infinite matter.
For all cases, the total energies per particle are almost zero at the energy minimum point,
which is largely underestimated because of the lack of short-range correlation in both the 0p0h and HF configurations.

We compare the results between three cases.
For $A=28$, we see the difference in the results between the present finite and infinite size calculations.
The finite size calculation reproduces the total energy of the infinite size one in the lower density region, but, overestimates it in the higher density region, which comes from the potential energy. 
The kinetic energy in finite size calculation reproduces that of infinite size one in overall density region.
On the other hand, the $A=132$ case reproduces the infinite number solutions for all of the Hamiltonian components in the overall density region.
This indicates that the 0p0h state with magic number $A=132$ ($N_g=33$) simulates well the properties of Fermi sphere of infinite nuclear matter.

\begin{figure}[b]
\begin{center}
\includegraphics[width=7.5cm,clip]{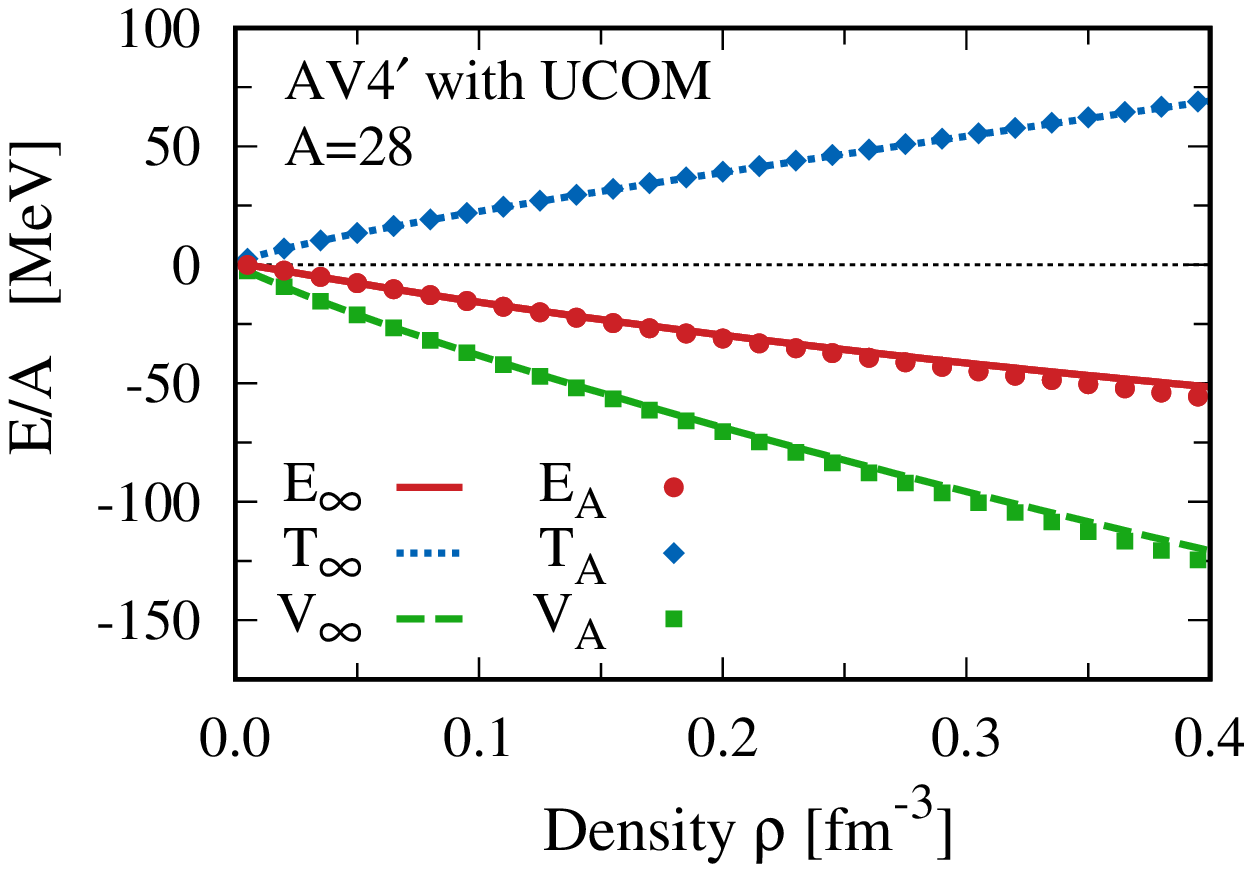}
\includegraphics[width=7.5cm,clip]{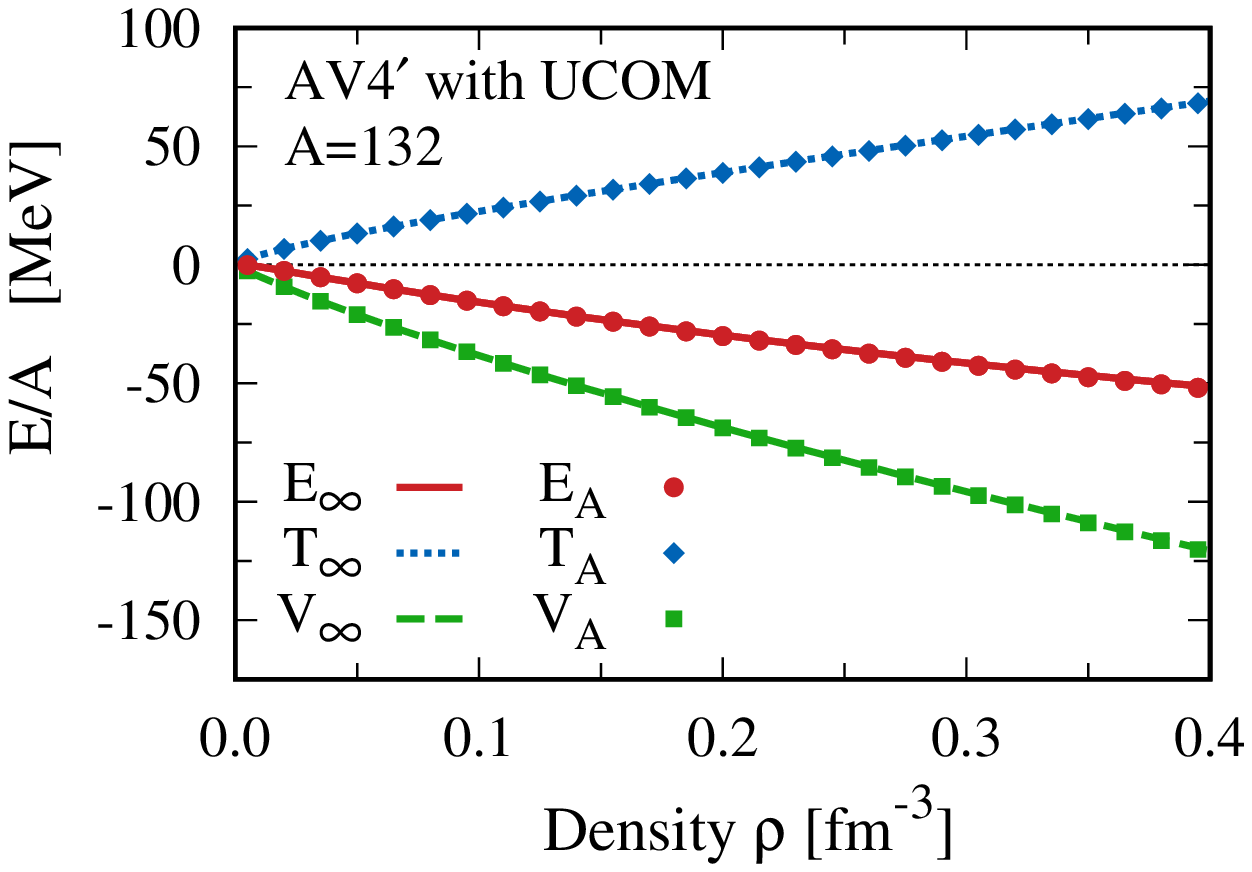}
\caption{Hamiltonian components per particle for the 0p0h state of symmetric nuclear matter with UCOM.
Mass numbers of $A=28$ (upper panel) and $A=132$ (lower panel). 
Notations are the same as those in Fig.~\ref{fig:HF_SNM}.}
\label{fig:HF+UCOM_SNM}
\end{center}
\end{figure}

Next, we see the effect of UCOM in nuclear matter with the 0p0h+UCOM wave function.
The UCOM introduces the short-range correlation as the 2p2h excitations in the wave function. It is interesting to investigate the validity of the present finite size approach in the correlated wave function as well as the previous 0p0h case.
We also calculate the infinite nuclear matter in the HF state with UCOM for comparison.

In Fig.~\ref{fig:HF+UCOM_SNM}, we show the energy per particle for the 0p0h configuration with central-UCOM as functions of density $\rho$.
In this case, all the solutions of $A=28, 132$ and infinite number show the negative total energies and this energy gain represents the effect of UCOM.
Among them, it is found that the $A=132$ case reproduces the infinite number solutions for every Hamiltonian components, similar to the 0p0h case.
This means that the present finite size approach can simulate the properties of the nuclear matter with infinite particle number in case of the correlated wave function.  

\begin{figure}[b]
\begin{center}
\includegraphics[width=7.5cm,clip]{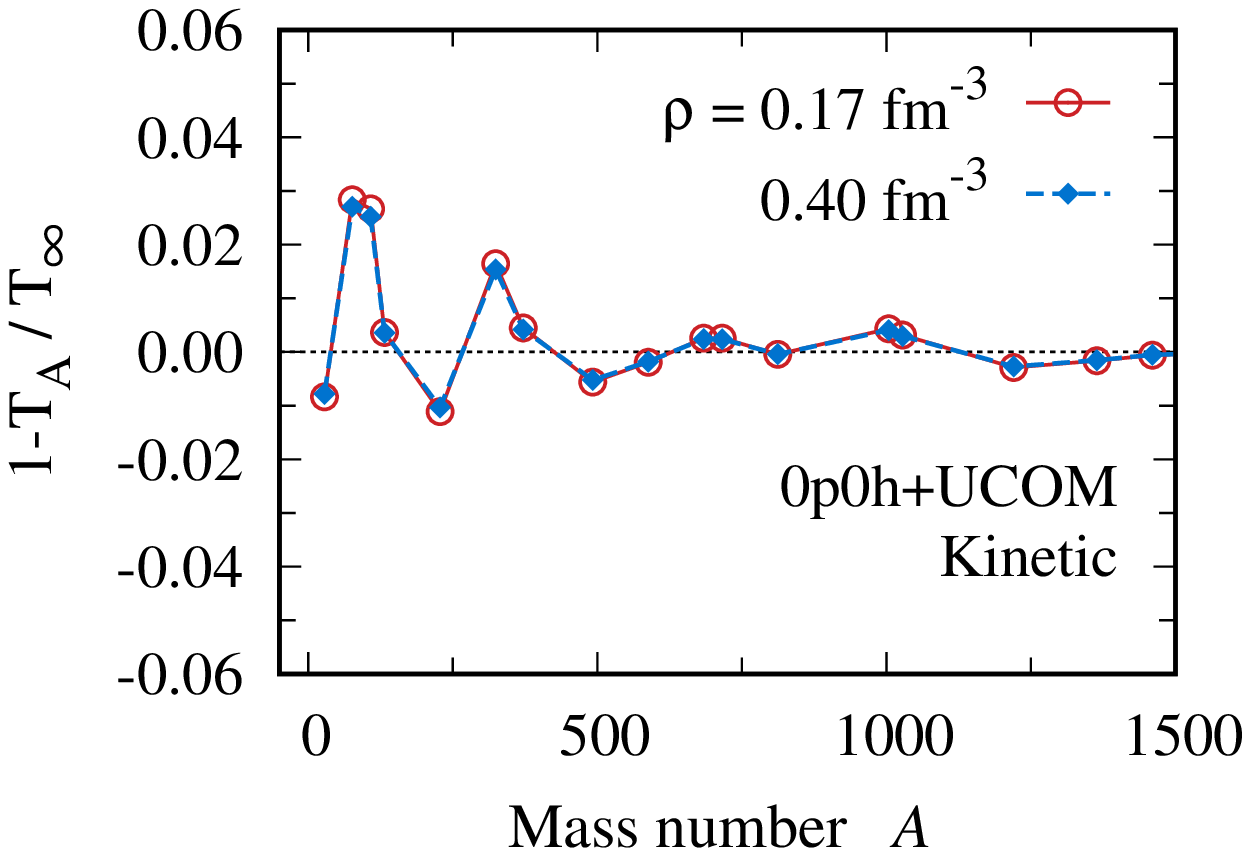}
\includegraphics[width=7.5cm,clip]{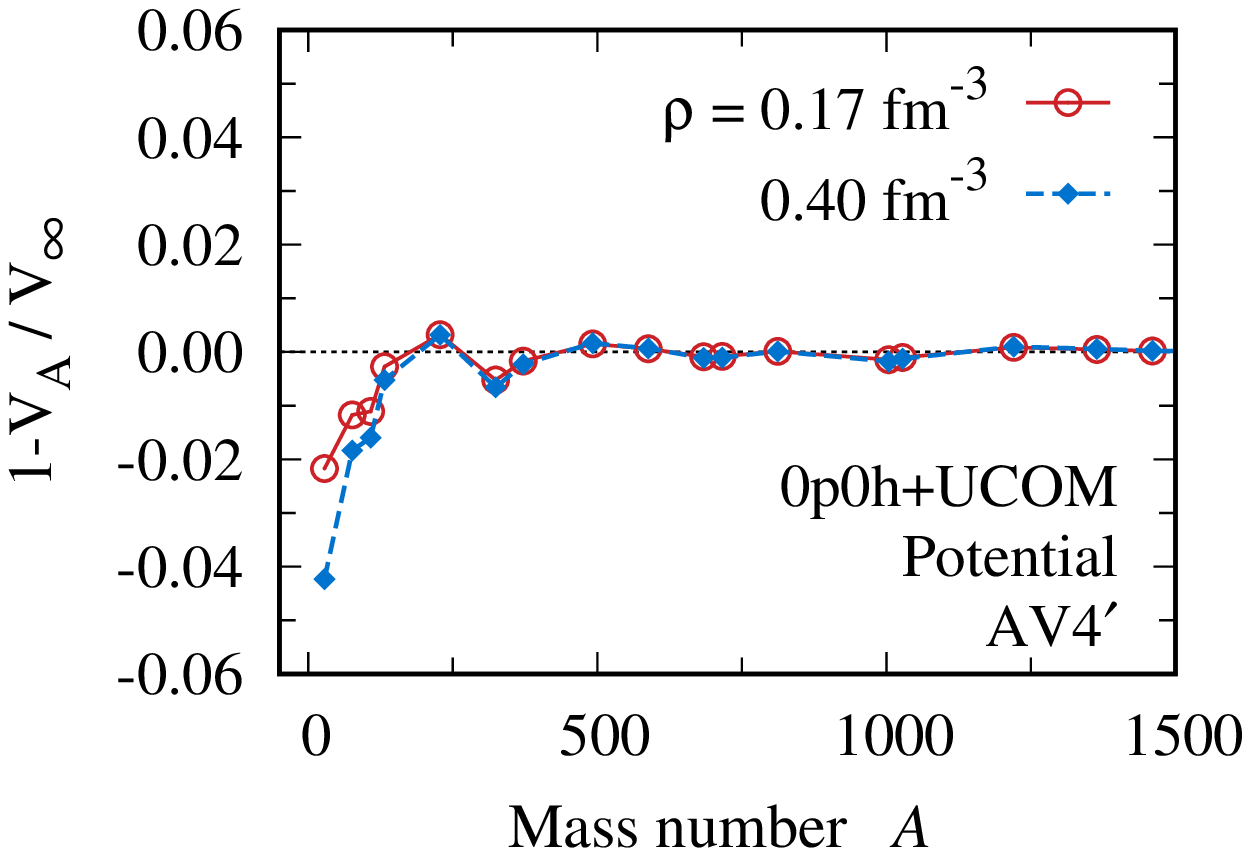}
\caption{Relative errors of the Hamiltonian components per particle in 0p0h+UCOM of symmetric nuclear matter with respect to the Hartree-Fock solutions of infinite matter with UCOM, as functions of mass number $A$. Kinetic (upper panel) and potential (lower panel) energies are shown with two kinds of densities $\rho$.}
\label{fig:A_depend}
\end{center}
\end{figure}

We examine the mass number dependence of the solutions in the 0p0h+UCOM wave function.
In Fig.~\ref{fig:A_depend}, we show the Hamiltonian components used in the 0p0h+UCOM wave function,
with respect to those of the infinite particle number with UCOM, as functions of mass number $A$ being magic numbers.
We also fix the density $\rho$ with two cases, normal (0.17 fm$^{-3}$) and higher (0.40 fm$^{-3}$) one.
In the results, as mass number $A$ increases, the relative errors of all kinds of Hamiltonian components rapidly decrease with less than 1$\%$ in both densities.
This results indicates that the individual Hamiltonian components converge to the values of infinite matter.
In smaller mass number region, it is found that the $A=132$ case, the fourth point from the left side, provides the good approximation for both of the kinetic and potential energies. 
This trend is commonly confirmed in other theories based on the finite size calculation \cite{gandolfi09,hagen14}.
Using this property, we choose $A=132$ throughout the present analysis of symmetric nuclear matter.

\begin{figure}[t]
\begin{center}
\includegraphics[width=7.5cm,clip]{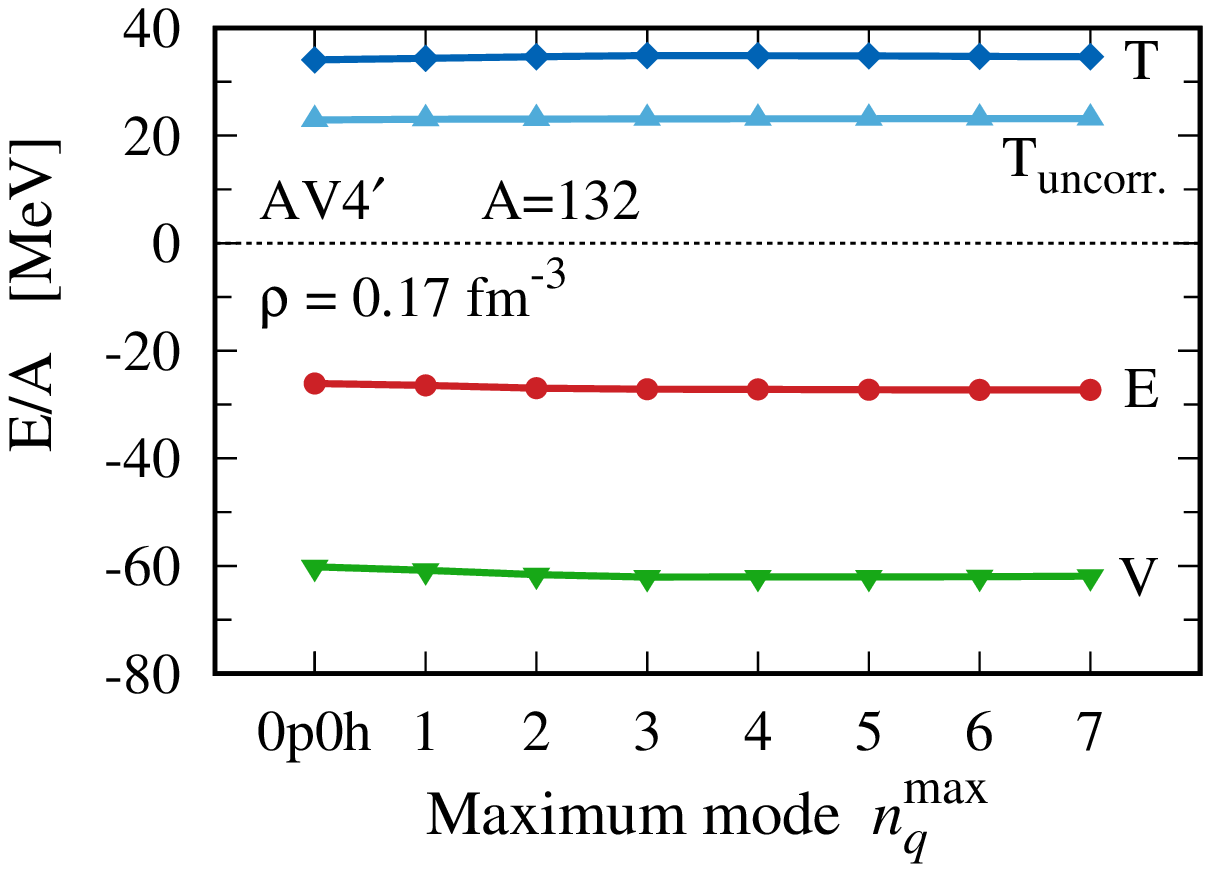}
\includegraphics[width=7.5cm,clip]{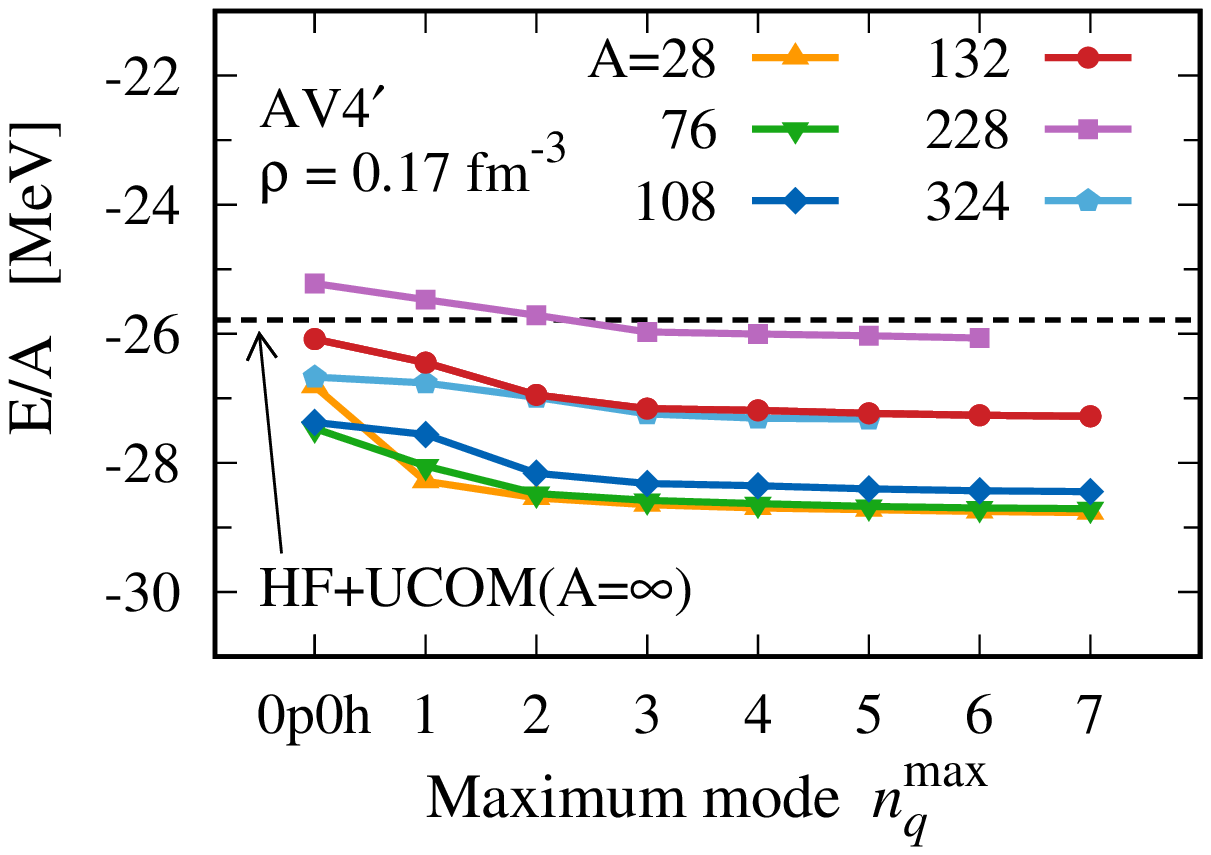}
\caption{Dependence of the solutions of symmetric nuclear matter on the maximum mode of transfer momentum $n_q^{\rm max}$ with several mass numbers in UCOM+HM.
Upper panel shows the case of $A=132$, where $E$, $T$, and $V$ are the total, kinetic and potential energies, respectively. The term of $T_{\rm uncorr.}$ is the uncorrelated one-body kinetic energy without UCOM part.
Lower panel shows the total energies of six kinds of mass numbers. Dashed line is the energy of the Hartree-Fock calculation with UCOM for infinite matter.}
\label{fig:SNM_qdep}
\end{center}
\end{figure}

We present our final results of symmetric nuclear matter with UCOM+HM including 2p2h configurations as high-momentum pairs.
In Fig.~\ref{fig:SNM_qdep}, we show the dependence of solutions on the maximum mode of transfer momentum $n_q^{\rm max}$,
which controls the space of the 2p2h configurations.
The density is fixed to be the normal value of $\rho=0.17$ fm$^{-3}$.
We increase $n_q^{\rm max}$ and confirm convergence of the solutions.
The energies at $n_q=0$ corresponds to the results with the 0p0h+UCOM wave function.
In the figures, the upper panel shows the Hamiltonian components in case of $A=132$.
The effect of high-momentum pairs is 1.2 MeV in total energy per particle and not large.
The kinetic energy slightly increases and potential energy gains as well.
This indicates that the UCOM can treat the dominant part of the correlations coming from the AV4$^\prime$  potential and that the 2p2h configurations describe the residual part.
We can confirm the UCOM effect on the kinetic energy by seeing the difference between total kinetic energy $T$ with UCOM and the uncorrelated part $T_{\rm uncorr.}$ without UCOM, where the latter is the expectation value of the one-body operator of $\sum_{i=1}^A t_i$ in Eq. (\ref{eq:ham}). This effect is about 12 MeV per particle.

\begin{figure}[t]
\begin{center}
\includegraphics[width=7.5cm,clip]{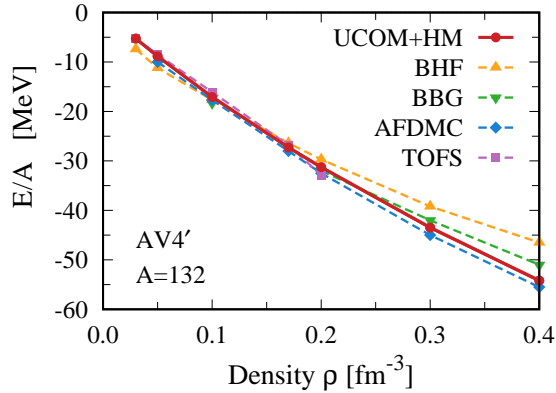}
\caption{Energies per particle of symmetric nuclear matter in UCOM+HM with mass number $A=132$ 
as functions of density $\rho$.
We also compare the results with other theories, the values of which are taken from Refs. \cite{baldo12,yamada18b}.}
\label{fig:SNM_cmp}
\end{center}
\end{figure}

\begin{table}[t]
\begin{center}
\caption{Hamiltonian components per particle in UCOM+HM for $A=132$ with $n_q^{\rm max}=6$ as functions of density $\rho$. Total, kinetic and potential energies are shown as $E$, $T$, $V$, respectively.
The units of energy and density are MeV/$A$ and fm$^{-3}$, respectively.
}
\label{tab:SNM_dns} 
\begin{tabular}{c|rrrrrrrrrr}
\noalign{\hrule height 0.5pt}
$\rho$~  &   0.03  & 0.05    &   0.10  &  0.17   & 0.20    & 0.30    & 0.40     \\
\noalign{\hrule height 0.5pt}
$E$~     & $-5.3$  & $-8.9$  & $-17.1$ & $-27.3$ & $-31.3$ & $-43.5$ & $-54.2$  \\
$T$~     & $~9.5$  & $13.8$  & $~23.2$ & $~34.7$ & $~39.4$ & $~54.3$ & $~68.7$  \\
$V$~     & $-14.8$ & $-22.7$ & $-40.2$ & $-62.0$ & $-70.7$ & $-97.7$ & $-122.9$ \\
\noalign{\hrule height 0.5pt}
\end{tabular}
\end{center}
\end{table}

In the lower panel of Fig.~\ref{fig:SNM_qdep}, we show the mass number dependence of the total energies per particle.
We show six cases of magic numbers $A$. 
At the level of 0p0h+UCOM, the $A=132$ calculation provides the total energy close to the value of the infinite number.
The converging energies of each mass number differ in the range of about 2.5 MeV and
the $A=132$ case provides the middle value among them.

Finally, in Fig.~\ref{fig:SNM_cmp}, we show the total energies per particle of symmetric nuclear matter in UCOM+HM with $A=132$ as functions of the density $\rho$. 
This is the nuclear equation of state and we compare the results with other theories.
The present results are consistent to those of other theories in the overall density region.
This ensures that our UCOM+HM framework is valid to describe the symmetric nuclear matter. 
In detail, the results agree with each other in the region lower than the normal density,
and there exist the differences among theories in the higher density region, 
although trend is very similar to each other.
The Hamiltonian components in UCOM+HM for each density are summarized in Table. \ref{tab:SNM_dns}.

\subsection{Neutron matter}\label{sec:NM}

\begin{figure}[b]
\begin{center}
\includegraphics[width=7.5cm,clip]{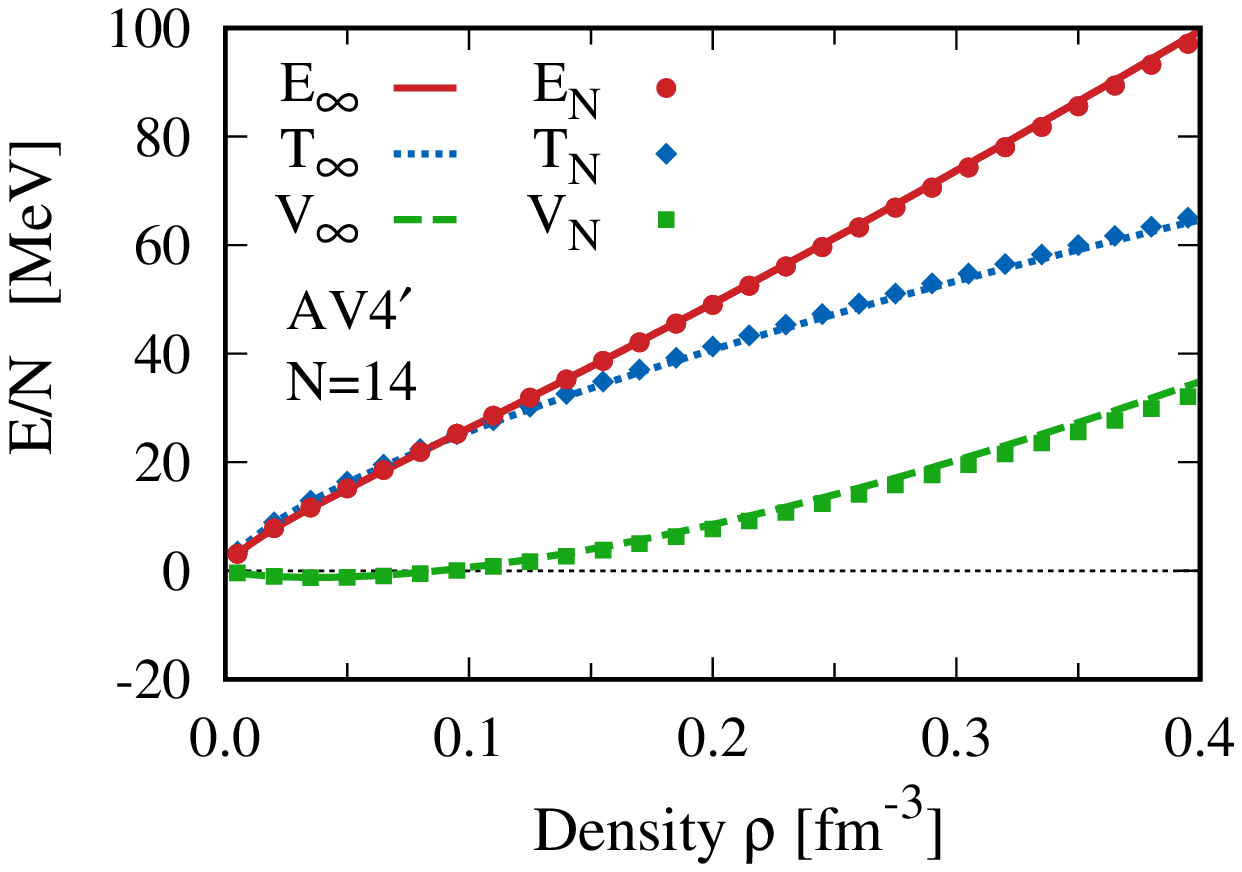}
\includegraphics[width=7.5cm,clip]{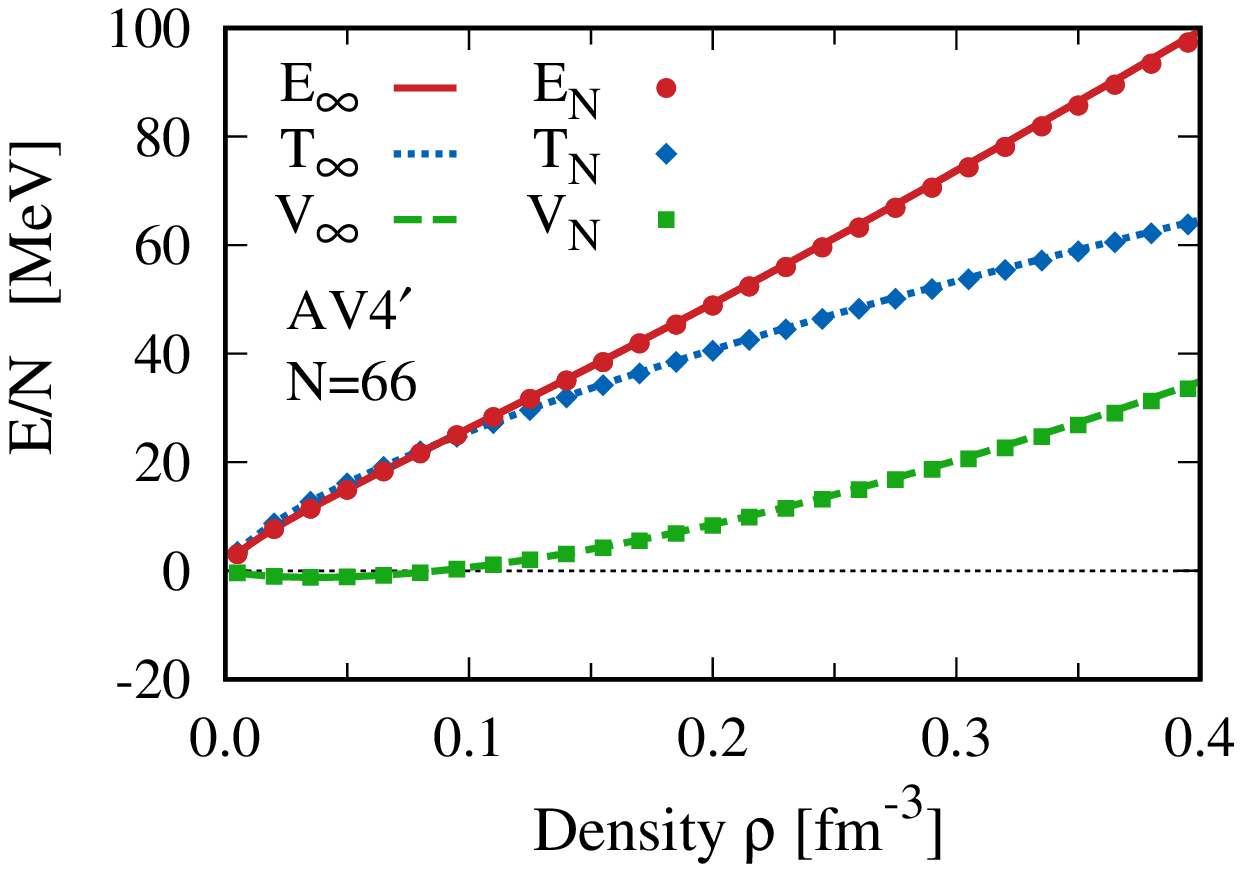}
\caption{Hamiltonian components per particle of neutron matter in the 0p0h configuration as functions of neutron density $\rho$ with dots in comparison with the Hartree-Fock solutions of infinite neutron matter with lines.
Neutron number is $N=14$ (upper panel) and $66$ (lower panel).
The values with red-circle, blue-diamond and green-square indicate the total ($E$), kinetic ($T$) and potential ($V$) energies, respectively.}
\label{fig:HF_NM}
\end{center}
\end{figure}

We investigate the neutron matter with UCOM+HM in a similar way of symmetric nuclear matter.
We start from the 0p0h configuration with finite neutron number, which corresponds to the Hartree-Fock state of the Fermi sphere for the infinite neutron matter.
In Fig.~\ref{fig:HF_NM}, we show the Hamiltonian components per particle as functions of neutron density $\rho$.
We show two cases of neutron magic number $N=14$ and $66$.
Both results reproduce the Hamiltonian components of the infinite neutron number in the overall density region.
In particular, even the $N=14$ case provides the good description rather than the case of $A=28$ for symmetric nuclear matter as shown in Fig.~\ref{fig:HF_SNM}.
This result may come from the property of the AV4$^\prime$ potential where the isospin $T=0$ channel is involved for symmetric nuclear matter, but not for neutron matter.
Results in Fig.~\ref{fig:HF_NM} indicate the reliability of the present approach with finite neutron numbers, similarly to the discussion on symmetric nuclear matter.

\begin{figure}[b]
\begin{center}
\includegraphics[width=7.5cm,clip]{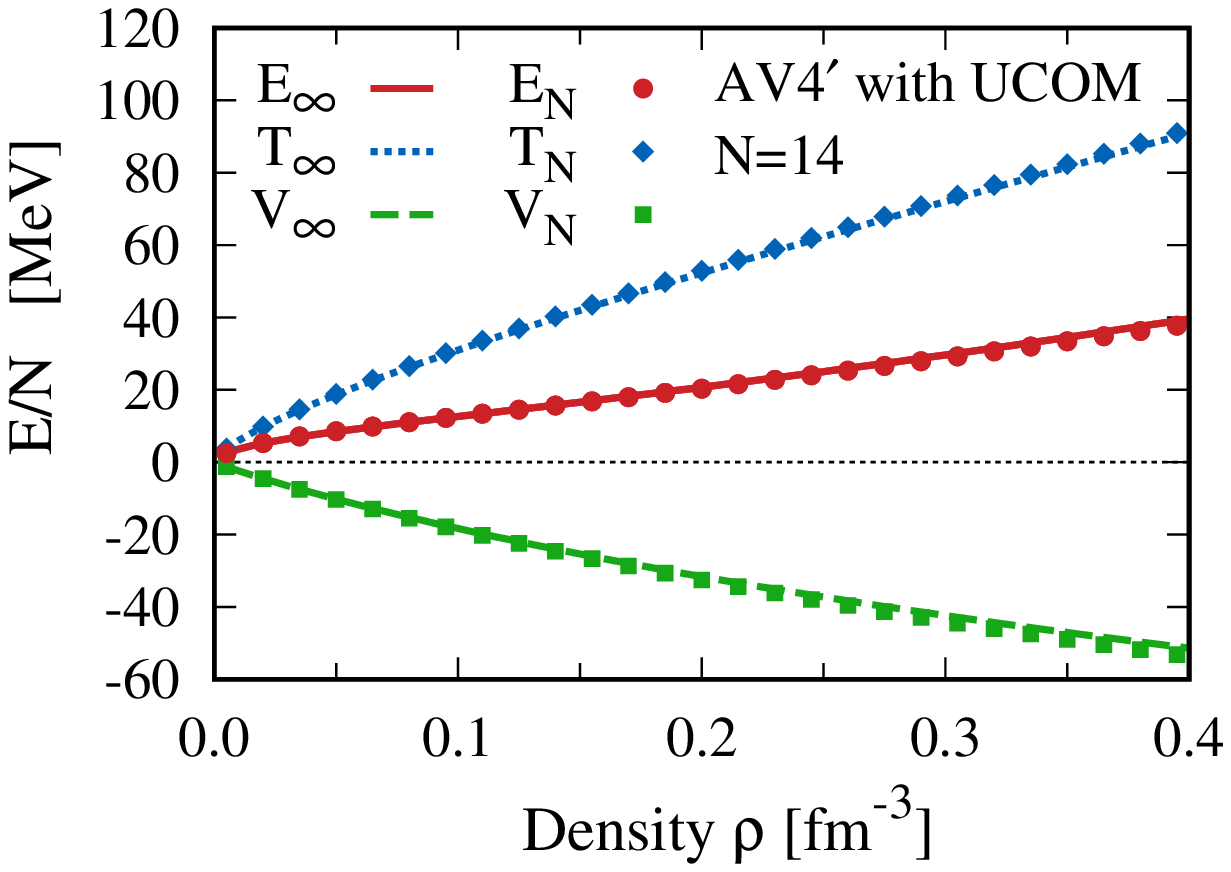}
\includegraphics[width=7.5cm,clip]{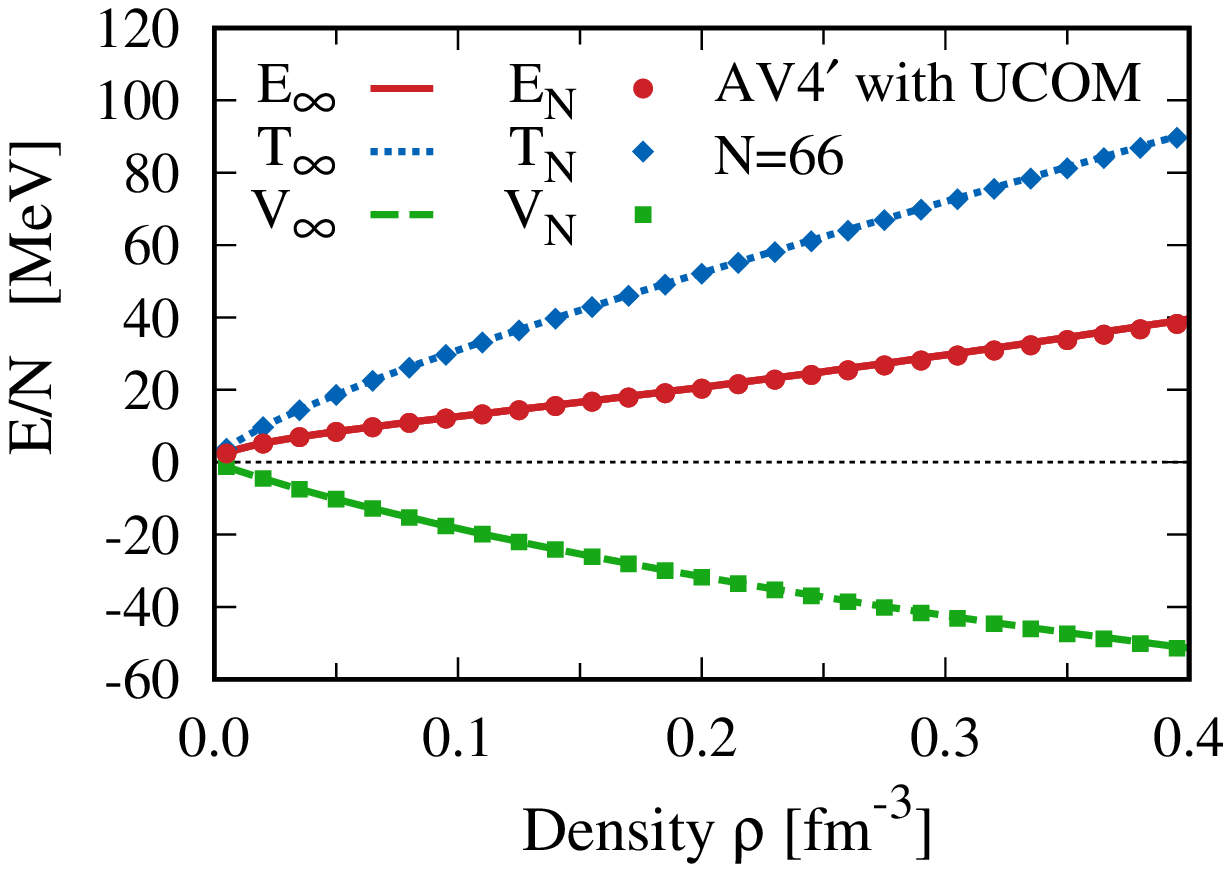}
\caption{Hamiltonian components per particle of neutron matter in the 0p0h+UCOM wave function for neutron numbers of $N=14$ (upper panel) and $N=66$ (lower panel). Notations are the same as those in Fig.~\ref{fig:HF_NM}.}
\label{fig:HF+UCOM_NM}
\end{center}
\end{figure}

Next, we discuss the 0p0h+UCOM wave function.
In Fig.~\ref{fig:HF+UCOM_NM}, we show the Hamiltonian components per particle of neutron matter as functions of neutron density $\rho$.
Because of the effect of UCOM for short-range correlation, 
the total and potential energies become lower, while kinetic energies increase due to the correlation.
In these results, it is also confirmed that the finite neutron number approach nicely reproduces the infinite number solutions with UCOM.

\begin{figure}[t]
\begin{center}
\includegraphics[width=7.5cm,clip]{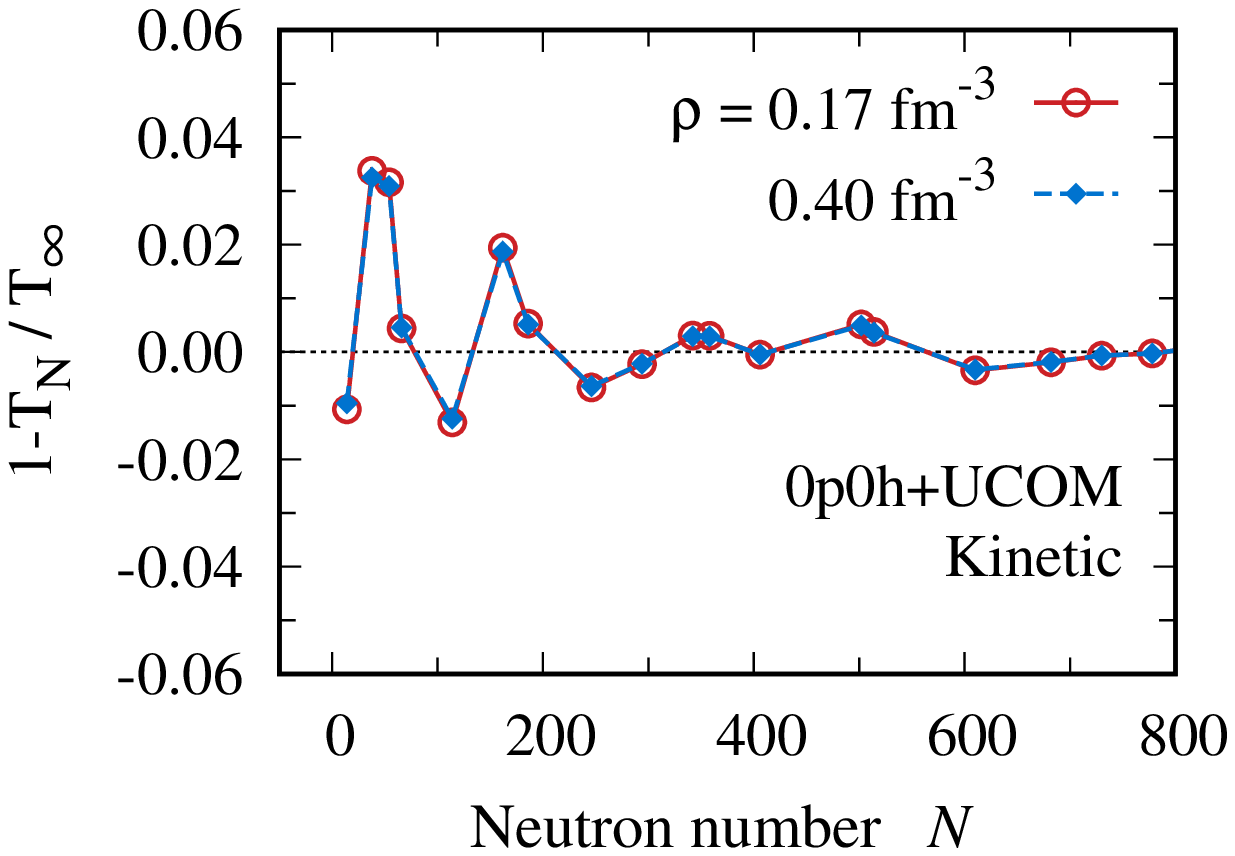}
\includegraphics[width=7.5cm,clip]{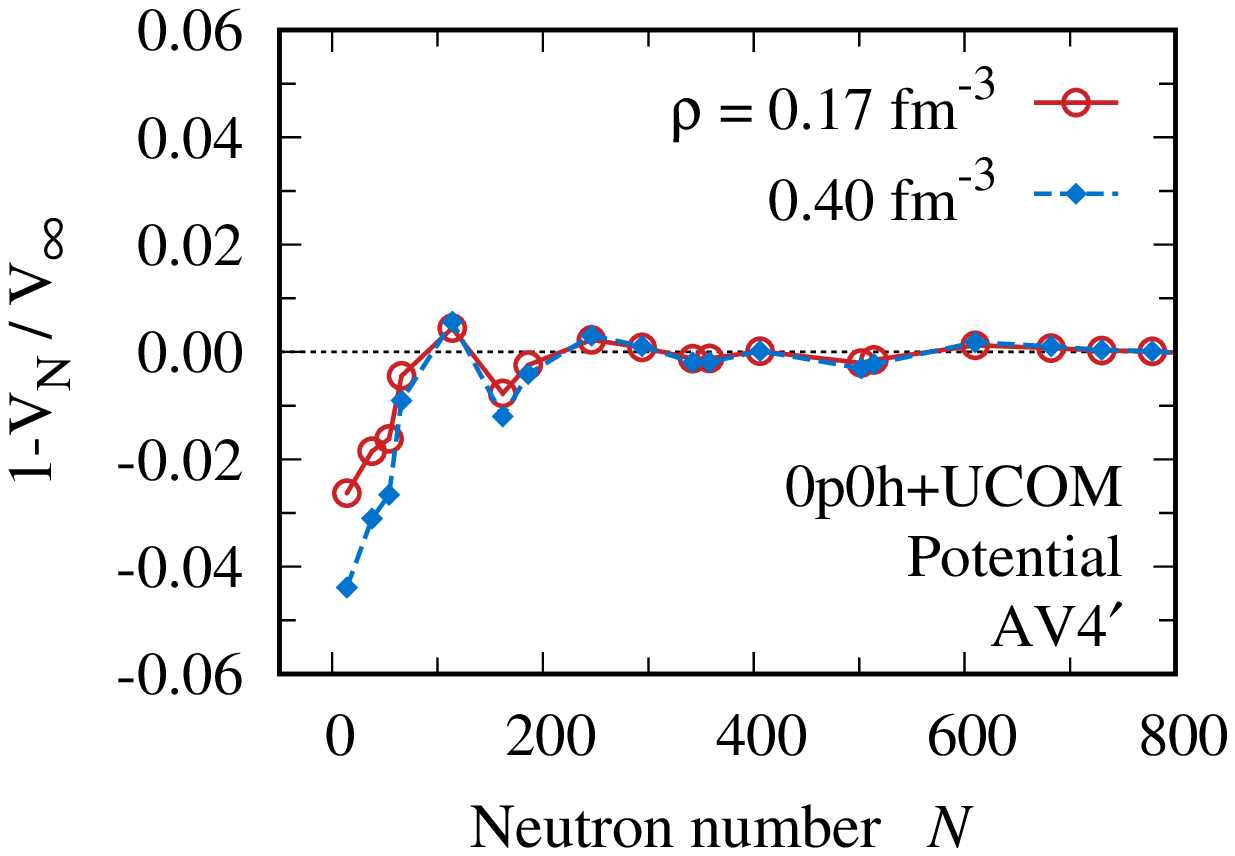}
\caption{Relative errors of the Hamiltonian components per particle in 0p0h+UCOM for neutron matter with respect to the Hartree-Fock solutions of infinite neutron matter with UCOM, as functions of neutron number $N$.
Kinetic (upper panel) and potential (lower panel) energies are shown with two kinds of neutron densities $\rho$.}
\label{fig:N_depend}
\end{center}
\end{figure}

We investigate the neutron number dependence of the solutions in the 0p0h+UCOM wave function.
In Fig.~\ref{fig:N_depend}, we show the Hamiltonian components per particle
with respect to those of the infinite neutron matter for two kinds of densities.
As a neutron number $N$ increases, the relative error of all kinds of Hamiltonian components reduces within $1\%$.
In the region of smaller neutron numbers, the $N=66$ case with a grid number of $N_g=33$, the fourth point from the left side, provides the good approximation for every component.
This trend is the same as that of symmetric nuclear matter as shown in Fig.~\ref{fig:A_depend}.
Considering these results, we adopt $N=66$ in the present analysis of neutron matter.

\begin{figure}[th]
\begin{center}
\includegraphics[width=7.5cm,clip]{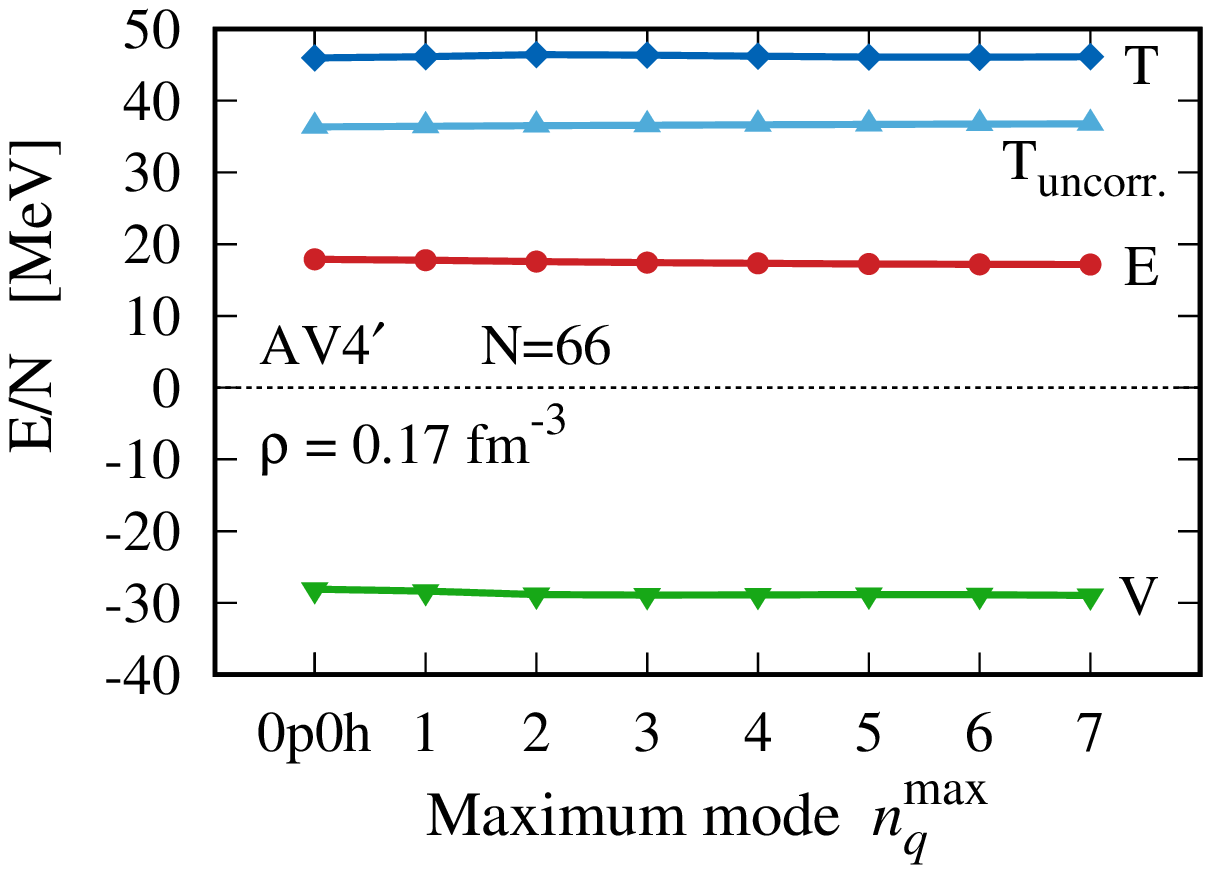}
\includegraphics[width=7.5cm,clip]{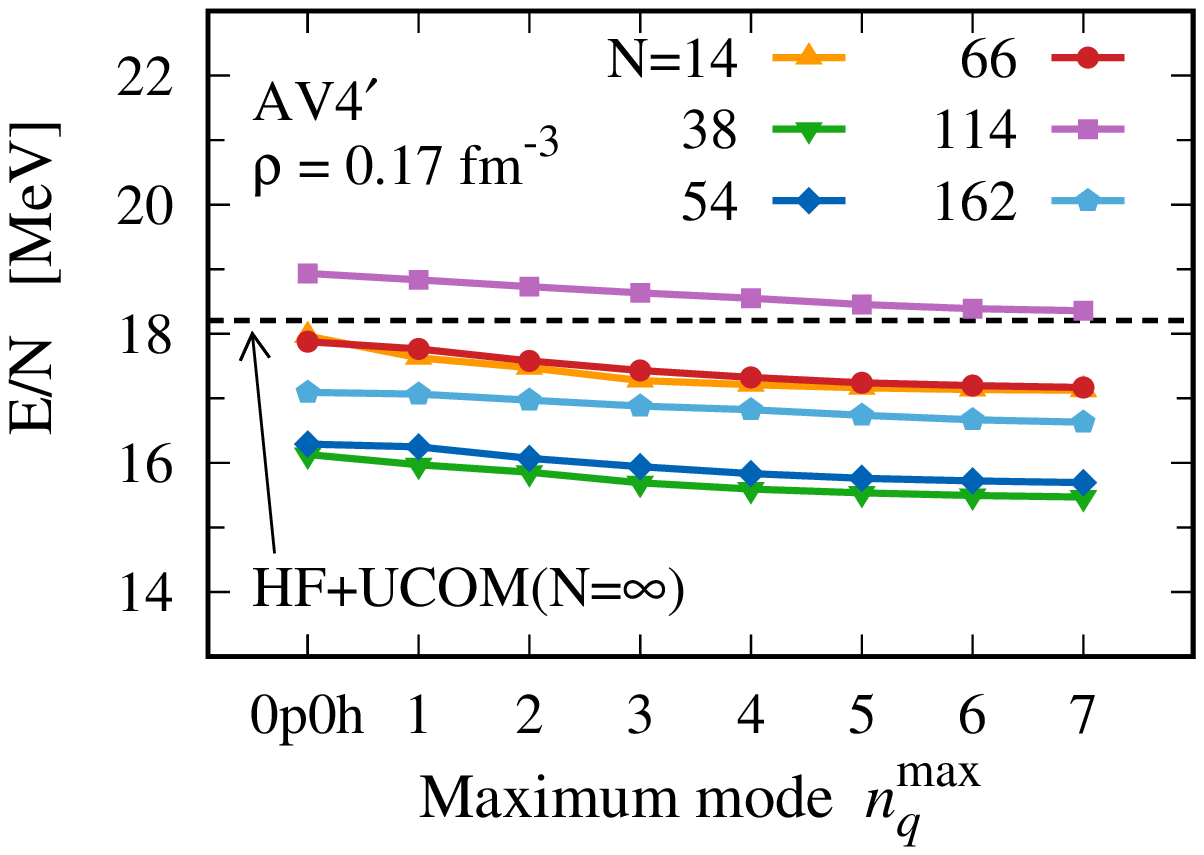}
\caption{Energies of neutron matter per particle increasing the maximum mode of transfer momentum $n_q^{\rm max}$ with several neutron numbers in UCOM+HM.
Upper panel shows the case of $N=66$, where $E$, $T$, and $V$ are the total, kinetic and potential energies, respectively. The term of $T_{\rm uncorr.}$ is the uncorrelated kinetic energy without UCOM.
Lower panel shows the total energies of six kinds of neutron magic numbers.
Dashed line is the total energy of the Hartree-Fock calculation with UCOM for infinite neutron matter.
}
\label{fig:NM_qdep}
\end{center}
\end{figure}

We add the 2p2h configurations into the 0p0h+UCOM wave function, which is the UCOM+HM wave function.
In Fig.~\ref{fig:NM_qdep}, we show the dependence of the maximum mode of transfer momentum $n_q^{\rm max}$,
which controls the space of the 2p2h configurations, on the solutions. 
We assume the neutron density $\rho=0.17$ fm$^{-3}$, the same value as used in the symmetric nuclear matter shown in Fig.~\ref{fig:SNM_qdep}.
We increase $n_q^{\rm max}$ until we get the convergence of the solutions.
In the upper panel, the Hamiltonian components per particle of the $N=66$ case are shown and convergence is clearly confirmed for every component. 
The effect of short-range correlation can be seen in the kinetic energy as the difference between the uncorrelated value $T_{\rm uncorr.}$ and total value $T$, which is about 9 MeV per particle.
In the lower panel, we show the neutron number dependence on the total energies.
The HF+UCOM energy in the infinite neutron number is shown as a reference.
We take six kinds of neutron magic numbers and their energy difference is within 3 MeV at converging points.
The $N=66$ cases provide almost the middle value among them.

\begin{figure}[t]
\begin{center}
\includegraphics[width=7.5cm,clip]{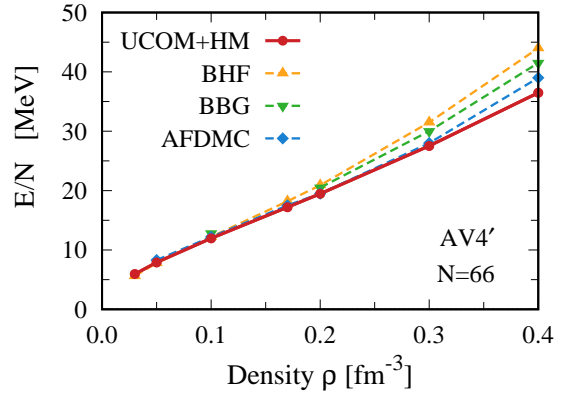}
\caption{Energies of neutron matter per particle in the present UCOM+HM method with a neutron number $N=66$ as functions of neutron density $\rho$. We compare the results with other theories, the values of which are taken from Ref. \cite{baldo12}.}
\label{fig:NM_cmp}
\end{center}
\end{figure}

Finally, in Fig.~\ref{fig:NM_cmp}, we show the total energies per particle in UCOM+HM as function of the neutron density $\rho$ in comparison with other theories.
We adopt $N=66$ and the present results are consistent to those of other theories in the overall density region.
It is found that the present UCOM+HM provides the similar energies to those of other theories in lower density region, while in the higher density region there are differences between theories including UCOM+HM.
The Hamiltonian components per particle of neutron matter in UCOM+HM for each neutron density are summarized in Table. \ref{tab:NM_dns}. 
It is noted that the density dependence of the present equation of state with AV4$^\prime$ potential is similar to those obtained with AV8$^\prime$ one \cite{gandolfi14}, in which two potentials have the same 
central component in the neutron-neutron channel.

\begin{table}[t]
\begin{center}
\caption{Hamiltonian components per particle in UCOM+HM for $N=66$ with $n_q^{\rm max}=6$ as functions of neutron density $\rho$. Total, kinetic and potential energies are shown as $E$, $T$, $V$, respectively.
The units of energy and density are MeV/$N$ and fm$^{-3}$, respectively.
}
\label{tab:NM_dns} 
\begin{tabular}{c|rrrrrrrrrr}
\noalign{\hrule height 0.5pt}
$\rho$~   &  0.03   & 0.05    &   0.10  &  0.17   & 0.20    & 0.30    & 0.40    \\
\noalign{\hrule height 0.5pt}
$E$~      & $~5.9$  & $~7.9$  & $~12.0$ & $~17.2$ & $~19.5$ & $~27.5$ & $~36.5$ \\
$T$~      & $13.4$  & $19.1$  & $~31.2$ & $~46.1$ & $~52.1$ & $~71.3$ & $~89.9$  \\
$V$~      & $-7.5$  & $-11.2$ & $-19.3$ & $-28.9$ & $-32.6$ & $-43.8$ & $-53.4$ \\
\noalign{\hrule height 0.5pt}
\end{tabular}
\end{center}
\end{table}

From all kinds of the results of symmetric nuclear and neutron matters in the UCOM+HM method, 
we can conclude that the present new variational method 
can be a powerful approach to investigate the nuclear matter property. 
In the present study, we focus on the short-range central correlation, which is nicely described in UCOM+HM.
We shall extend this approach to treat the non-central tensor and $LS$ forces in the bare $NN$ interaction and also the three-nucleon forces explicitly.
It is expected that the high-momentum components in the 2p2h configurations in UCOM+HM become important in the description of the tensor correlation for nuclear matter.

\section{Summary}\label{sec:summary}

We propose a new variational framework to describe the nuclear matter using the unitary correlation operator method (UCOM) for central correlation and the 2p2h excitations.
We describe the 2p2h excitations in terms of the transfer momentum in a nucleon pair, which changes the momentum of each nucleon to the opposite direction.
This transfer momentum can make a large relative momentum in a nucleon pair, regarding as a ``high-momentum pair'' (HM).
We name this new method ``UCOM+HM'', in which we can treat the high-momentum components in nuclear matter coming from the nucleon-nucleon interaction explicitly. 
We determine the central correlation functions in UCOM variationally and also increase the transfer momentum in the 2p2h configurations until we get the convergence of the solutions of nuclear matter.

In order to include the 2p2h effect in nuclear matter, we take the finite size approach with finite particle numbers.
We choose the particle numbers as magic numbers on a lattice in a discretized momentum space under the periodic boundary condition.
We investigate the particle number dependence of the present solutions, which provides an energy difference within a few MeV per particle.
Among various particle numbers, the numbers of $A=132$ and $N=66$ are preferable to simulate the properties of infinite matter for symmetric nuclear and neutron matters, respectively, the results of which are common with other theories. 
 
We show the validity of the present UCOM+HM method to describe the short-range correlation in nuclear matter using the AV4$^\prime$ central potential with short-range repulsion.
The UCOM nicely works to describe the short-range correlation in nuclear matter as well as in finite nuclei.
The additional 2p2h excitations make the energy gain by about few MeV per particle at around normal nuclear density.
As a result, the present method provides the nuclear equations of state for symmetric nuclear and neutron matters,
which are consistent to those in other theories, such as Brueckner-Hartree-Fock and auxiliary field diffusion Monte Carlo, in the overall density region. 

Based on the success of the present UCOM+HM method, in the next step, we employ the bare $NN$ interaction such as the AV8$^\prime$ potential having tensor and $LS$ forces and investigate the nuclear equation of state realistically.
In particular, the tensor correlation can be treated in terms of the 2p2h excitations with a wide momentum space in the present method.
The inclusion of the three-nucleon force is also an interesting subject for the quantitative description of the nuclear equation of state.

\section*{Acknowledgments}
This work was supported by JSPS KAKENHI Grants No. JP18K03660 and No. JP16K05351,
and the National Natural Science Foundation of China (Grants No. 11822503, No. 11575082).

\section*{References}
\def\JL#1#2#3#4{ {{\rm #1}} \textbf{#2}, #4 (#3)}  
\nc{\PR}[3]     {\JL{Phys. Rev.}{#1}{#2}{#3}}
\nc{\PRC}[3]    {\JL{Phys. Rev.~C}{#1}{#2}{#3}}
\nc{\PRA}[3]    {\JL{Phys. Rev.~A}{#1}{#2}{#3}}
\nc{\PRE}[3]    {\JL{Phys. Rev.~E}{#1}{#2}{#3}}
\nc{\PRL}[3]    {\JL{Phys. Rev. Lett.}{#1}{#2}{#3}}
\nc{\NP}[3]     {\JL{Nucl. Phys.}{#1}{#2}{#3}}
\nc{\NPA}[3]    {\JL{Nucl. Phys.}{A#1}{#2}{#3}}
\nc{\PL}[3]     {\JL{Phys. Lett.}{#1}{#2}{#3}}
\nc{\PLB}[3]    {\JL{Phys. Lett.~B}{#1}{#2}{#3}}
\nc{\PTP}[3]    {\JL{Prog. Theor. Phys.}{#1}{#2}{#3}}
\nc{\PTPS}[3]   {\JL{Prog. Theor. Phys. Suppl.}{#1}{#2}{#3}}
\nc{\PTEP}[3]   {\JL{Prog. Theor. Exp. Phys.}{#1}{#2}{#3}}
\nc{\PRep}[3]   {\JL{Phys. Rep.}{#1}{#2}{#3}}
\nc{\PPNP}[3]   {\JL{Prog.\ Part.\ Nucl.\ Phys.}{#1}{#2}{#3}}
\nc{\JP}[3]     {\JL{J. of Phys.}{#1}{#2}{#3}}
\nc{\EPJA}[3]   {\JL{Eur. Phys. J. A}{#1}{#2}{#3}}
\nc{\andvol}[3] {{\it ibid.}\JL{}{#1}{#2}{#3}}

\end{document}